\documentclass[a4paper]{article}
\usepackage{threeparttable}
\usepackage{stmaryrd}
\usepackage{booktabs}
\usepackage{amssymb}
\usepackage{mathrsfs}
\usepackage{latexsym}
\usepackage{epsfig}
\usepackage{subfigure}
\usepackage{graphicx}
\usepackage{epstopdf}
\usepackage{amsthm}
\usepackage{amsfonts}
\usepackage{amsmath}
\usepackage{amssymb,amsmath,amsthm,amscd}
\usepackage{algorithm}
\usepackage{algorithmic}
\usepackage{multirow}
\usepackage{xcolor}
\usepackage{array}
\usepackage{bm}
\usepackage[numbers,sort&compress]{natbib}
\usepackage{lscape}
\usepackage[indentafter]{titlesec}
\usepackage[colorlinks,linkcolor=red,linkcolor=blue,linktocpage]{hyperref}
\usepackage{threeparttable}
\usepackage{caption}
\usepackage[top=1 in,bottom=1 in,left=1.0 in,right=1.0 in]{geometry}
\usepackage{enumitem,calc}

\newcommand\preitem{\mdseries\textbullet\space}
\newlist{desclist}{description}{3}
\setlist[desclist,1]{format=\preitem\bfseries,leftmargin=\widthof{\preitem},style=sameline}
 \makeatletter

\newcommand{\Rmnum}[1]{\expandafter\@slowromancap\romannumeral #1@}
\floatname{algorithm}{Method}
\newtheorem{theorem}{Theorem}%[section]

\newtheorem{remark}{Remark}%[section]

\numberwithin{equation}{section}
\hypersetup{colorlinks=true, citecolor=blue}

\def \Re {\mathrm{Re}}
\def \Im {\mathrm{Im}}

\newcommand*{\caso}[1]{\overline{#1}}

\title{The  integrable semi-discrete nonlinear Schr\"odinger equations with nonzero backgrounds:
Bilinearization-reduction approach}

\author{Xiao Deng$^1$,~~Kui Chen$^2$,~~Hongyang Chen$^1$,
~~Da-jun Zhang$^{3,4}$\footnote{Corresponding author. Email: djzhang@staff.shu.edu.cn}, \\
{\small $^1$Research Center for Data Hub and Security, Zhejiang Lab, Hangzhou, Zhejiang 311100,  China}\\
{\small $^2$Research Center for Frontier Fundamental Studies, Zhejiang Lab, Hangzhou, Zhejiang 311100, China}\\
{\small $^3$Department of Mathematics, Shanghai University, Shanghai 200444, China}\\
{\small $^{4}$Newtouch Center for Mathematics of Shanghai University,  Shanghai 200444, China}}

\date{\today}
\begin{document}
\maketitle

\begin{abstract}
In this paper the classical and nonlocal semi-discrete nonlinear  Schr\"{o}dinger (sdNLS) equations
with nonzero backgrounds are solved by means of the bilinearization-reduction approach.
In the first step of this approach, the unreduced sdNLS system  with a nonzero background
is bilinearized and its solutions are presented in terms of quasi double Casoratians.
Then, reduction techniques are implemented to deal with complex and nonlocal reductions,
which yields solutions for the four classical and nonlocal sdNLS equations
with a plane wave background or a hyperbolic function background.
These solutions are expressed with explicit formulae and allow classifications
according to canonical forms of certain spectral matrix.
In particular, we present explicit  formulae for general rogue waves for the classical focusing sdNLS equation.
Some obtained solutions are analyzed and illustrated.

\begin{description}
\item[Keywords:] semi-discrete nonlinear Schr\"{o}dinger equation,  nonlocal, bilinear, reduction,
  double Casoratian, nonzero background.
\end{description}
\end{abstract}

\tableofcontents

%\newpage
\vskip 25pt

\section{Introduction}\label{sec-1}

The purpose of this paper is
to present and classify solutions with nonzero backgrounds
for the  integrable semi-discrete  nonlinear  Schr\"{o}dinger (sdNLS) equation
\begin{equation}\label{sdnls}
    i\partial_t Q_n=Q_{n+1}-2Q_n+Q_{n-1} -\delta |Q_n|^2(Q_{n+1}+Q_{n-1}), ~~ ~~ (\delta=-1)
\end{equation}
and its various nonlocal analogues (see \eqref{sec2-dNLS}).
Here $i$ is the imaginary unit, $Q_n=Q(n,t)$ is a complex function of
$(n,t)\in \mathbb{Z}\times \mathbb{R}$, $|Q_n|^2=Q_nQ_n^*$
and $*$ stands for complex conjugate.
Equation \eqref{sdnls} is also known as the Ablowitz-Ladik (AL) equation
since it is first presented by  Ablowitz and Ladik in \cite{AL-JMP-1976} in 1976 as an integrable discretization
of the (continuous) nonlinear  Schr\"{o}dinger (NLS) equation
\begin{equation}\label{nls}
i q_t=q_{xx}-\delta |q|^2q.
\end{equation}
Note that the above equation is called focusing NLS and defocusing NLS equations when $\delta=-1$ and $1$ respectively.

There is another semi-discrete NLS,
\begin{equation}\label{sdnls-n}
    i\partial_t Q_n=Q_{n+1}-2Q_n+Q_{n-1} -\delta |Q_n|^2Q_{n},
\end{equation}
which is not integrable but more significant in physics.
It serves as a model of optical discrete spatial solitons in nonlinear waveguide arrays,
which was first theoretically predicted in 1988 by Christodoulides and Joseph \cite{CJ-OL-1988},
and first realized experimentally by Eisenberg et al in 1998 \cite{ESMBA-PRL-1998}
in an one-dimensional identical infinite waveguide array with optical Kerr effect,
and later realized in more experiments, e.g. \cite{CLS-Nat-2003}.
There are many review papers and books about optical discrete solitons based on equation \eqref{sdnls-n},
which can be referred for readers to, e.g.
\cite{LSCASS-PR-2008,FW-PR-1998,FG-PR-2008,P-book-2009,YKMT-RMP-2011}
and references therein.
Besides,
equation \eqref{sdnls-n} desribes solitons in crystals due to dislocations.
It appears as a proximation of the one-dimensional Frenkel-Kontorova model \cite{BK-PR-1998}.
It was also derived from an anharmonic interatomic interaction chain \cite{CKKS-PRB-1993},
the so-called Fermi-Pasta-Ulam (FPU) model.
In addition to physics,
biologically, equation \eqref{sdnls-n} is used
to model energy transfer (in the form of vibration solitons) along $\alpha$-helical protein molecules
in Davydov's theory \cite{D-JTB-1977}.
Equation \eqref{sdnls-n} was derived from Davydov's Hamiltonian
in \cite{ST-PTRSA-1985} (also see \cite{S-PR-1992})
under an assumption of the average values of the longitudinal displacement of an amino acid being independent of time.
For more aspects of the nonintegrable sdNLS equation \eqref{sdnls-n},
one may also refer to \cite{K-Book-2009}.

Compared with \eqref{sdnls-n}, the sdNLS equation \eqref{sdnls}
is not as significant in application as \eqref{sdnls-n},
but it is still quite interesting in both physics and mathematics.
Since \eqref{sdnls} is integrable,
people studied \eqref{sdnls-n} as a perturbation of \eqref{sdnls} \cite{CKKS-PRB-1993,BK-PR-1998}.
In addition, the sdNLS equation \eqref{sdnls} is connected with
a Heisenberg lattice with a gauge transformation \cite{I-JNMP-1982}.
The equation is also linked to the Toda lattice \cite{AL-JMP-1976}.
A detailed description of the link can be found in Appendix C of \cite{APT-book-2004}.
Geometrically, the sdNLS equation \eqref{sdnls} describes solitons
along vortex  filament (cf.\cite{H-JFM-1972} in the continuous case),
which has been formulated from the motion of discrete curve \cite{DS-JMP-1995,HW-PLA-1996}
and from the motion of discrete surface \cite{HW-JPSJ-1996} as well.

The integrable sdNLS equation \eqref{sdnls} is exactly solvable.
Speaking of its solutions with nonzero background, we means
those solutions $Q_n$ which do not tend to zero when  $|n| \to \infty$.
Various methods have been employed to solve \eqref{sdnls} (focusing or defocusing case)
with  nonzero backgrounds,
for example, the inverse scattering transform (IST)
\cite{VK-IP-1992,ABP-IP-2007,vdM-JNMP-2015,P-JMP-2016,PV-SAPM-2016},
Hirota's bilinear method \cite{N-JPSJ-1991},
Kadomtsev-Petviashvili (KP)-reduction from solutions of the 2-dimensional Toda equation \cite{MO-JPSJ-2006,OY-JPA-2014},
Darboux transformation \cite{WY-JMP-2018},
and a special ansatz for elliptic solutions \cite{CCX-PLA-2006}.
Most of these researches are based on the plane wave  nonzero background (of exponential type).
Note that the exponential plane wave  leads to breathers
\cite{K-SPD-1977,KI-JPSJ-1977,KI-JPSJ-1978,M-SAPM-1979}
and rogue waves \cite{P-JAMSB-1983}  for the continuous NLS equation \eqref{nls} of focusing case,
and the envelope $|q|$ lives on a nonzero constant background.
For the sdNLS equation \eqref{sdnls} with a plane wave background,
its solutions are typically breathers and rogue waves as well
\cite{P-JMP-2016,WY-JMP-2018,AAL-PRE-2011,ADUCA-JO-2013}.
It is well known that rogue wave solutions can be obtained by taking certain limits from
breathers, which has been demonstrated for the sdNLS equation \eqref{sdnls}
in \cite{P-JMP-2016,WY-JMP-2018,AAS-PRE-2010}.
A remarkable result for explicit determinant expression for a general high order rogue
wave solution was given by Ohta and Yang in \cite{OY-JPA-2014}.
In addition, the sdNLS equation \eqref{sdnls} also admits rogue waves living on
an elliptic function background \cite{CP-SAPM-2024}.
Note that both discrete breathers and discrete rogue waves are
physically significant \cite{FG-PR-2008,BKA-OL-2009,EY-PLA-2015,HCFK-PLA-2018,MCKY-PRE-2022}.

In this paper, our aims are mainly on  deriving solutions with a plane wave background
for the focusing sdNLS equation \eqref{sdnls} ($\delta=-1$).
The obtained solutions will be breathers, rogue waves and their combination.
Our means is the so-called
bilinearization-reduction (B-R) approach, which has proved effective in finding solutions
for those equations that involve complex reductions.
Take the focusing NLS equation \eqref{nls} $(\delta=-1)$ as an example,
which is obtained from the coupled system
\begin{subequations}\label{nls-c}
\begin{align}
& i q_t=q_{xx}-q^2 r,\\
& i r_t =-r_{xx} +q r^2
\end{align}
\end{subequations}
through a reduction $r=-q^*$.
In the B-R approach, we first solve the unreduced system \eqref{nls-c} using bilinear
method and obtain solutions of the bilinear equation in terms of double Wronskians.
At this stage there is no complex reduction involved.
Then, impose constraints on the two generating vectors of the double Wronskians
so that the reduction $r=-q^*$ is satisfied.
In principle, constraint conditions can boil down to  some matrix equations
of which the solutions lead to classification of the solutions of the reduced equation \eqref{nls}.
For more details one may refer to review papers \cite{Z-book-2020,Z-APS-2023}.
The  B-R approach was introduced in 2018 \cite{CZ-AML-2018,CDLZ-SAPM-2018,DLZ-AMC-2018}
for solving nonlocal integrable systems.
Nonlocal integrable systems were first systematically proposed
by Ablowitz and Musslimani in 2013 \cite{AM-PRL-2013}.
In their settings, the reduction $r=\delta q^*$ for the unreduced NLS system \eqref{nls-c}
is replaced by $r(x,t)=\delta q^*(-x,t)$, and the reduced equation, i.e.
\begin{equation}\label{nls-n}
i q_t(x,t)=q_{xx}(x,t)-\delta q^2(x,t)q^*(-x,t),
\end{equation}
is still integrable but nonlocal in space.
Nonlocal integrable systems have drawn intensive attention after Ablowitz-Musslimani's pioneer work \cite{AM-PRL-2013},
for example,
\cite{AM-PRE-2014,SMM-PRE-2014,MZ-AML-2014,Y-AML-2015,F-Nonl-2016,AM-Nonl-2016,GS-JMP-2017,C-SAPM-2018,
YY-SAPM-2018,Z-SAPM-2018,AFLM-SAPM-2018,Y-PRE-2018,L-SAPM-2019,AM-JPA-2019,FZ-ROMP-2019,
L-CTP-2020,ALM-Nonl-2020,RCPMH-PD-2020,GPZ-PLA-2020,RS-JDE-2021,RS-CMP-2021}.
It has been a common understanding that solving nonlocal integrable systems
is essentially implementing reduction techniques.
The B-R approach provides a bilinear approach not only to nonlocal integrable systems but also to
classical ones. It has been successfully applied to various continuous equations, e.g.
\cite{SSZ-ND-2019,CLZ-AML-2019,LWZ-ROMP-2020,WWZ-2020,
W-ND-2021,LWZ-ROMP-2022,LWZ-SAPM-2022,WW-CNSNS-2022,WW-ND-2022,WWZ-CPB-2022,SLA-JPA-2023}
as well as to semi-discrete \cite{DLZ-AMC-2018,SWZ-AML-2021} and fully discrete ones \cite{ZFF-arX-2024}.
A recent progress is extending the B-R approach to the focusing NLS equation \eqref{nls}
with nonzero backgrounds \cite{ZLD-OCNMP-2023}.
As a result, explicit breathers and rogue waves in double Wronskian form were obtained.
Rogue waves are rational solutions, which, in principle, can be obtained from certain limit procedure.
For the focusing NLS equation, its high order rogue wave solutions in  special determinant forms
have been obtained in 1986 \cite{EKK-SPD-1986}
and then much later in 2012 \cite{GLL-PRE-2012} from Darboux transformation (together with limit procedures)
and in the same year from the KP-reduction approach \cite{OY-PRSA-2012}.
For the sdNLS equation \eqref{sdnls}, its high order  rogue waves in determinant forms
have been obtained in \cite{OY-JPA-2014} and \cite{WY-JMP-2018}
from KP-reduction and Darboux transformation, respectively.
In this paper, we will develop the B-R reduction approach
to solve the sdNLS equation \eqref{sdnls} and its nonlocal analogues (see \eqref{sec2-dNLS}) with nonzero backgrounds,
including a plane wave background and a hyperbolic function background.
We will see that in this approach we can not only  obtain  explicit breather and rogue wave solutions
in quasi double Casoratian form, but also classify solutions according to the canonical forms of certain matrix.

The paper is organized as follows.
In Sec.\ref{sec-2}  we present the unreduced sdNLS system, its Lax pair and classical and nonlocal reductions.
In Sec.\ref{sec-3} the unreduced sdNLS system is bilinearized and then solved
with solutions given in terms of quasi double Casoratians.
Sec.\ref{sec-4} displays the reduction techniques, which gives
explicit  solutions for the reduced equations.
Then, dynamics of some obtained solutions (including rogue waves) are analyzed and illustrated in Sec.\ref{sec-5}.
The final section devotes to conclusions.
There is an appendix which provides a proof for Theorem \ref{Th-1}.

\section{Unreduced, classical and nonlocal sdNLS equations}\label{sec-2}

For the sdNLS equation \eqref{sdnls}, the corresponding unreduced system reads
\begin{subequations}\label{sec2-AL2}
\begin{eqnarray}
    i\partial_t Q_{n}=(1-Q_nR_n)(Q_{n+1}+Q_{n-1})-2Q_n,\label{sec2-AL2-equ1}\\
    -i\partial_t R_{n}=(1-Q_nR_n)(R_{n+1}+R_{n-1})-2R_n,
\end{eqnarray}
\end{subequations}
which has a Lax pair \cite{AL-JMP-1976}
\begin{subequations}\label{Lax}
\begin{align}\label{sec2-spectral}
  &      \Theta_{n+1}=\mathcal M_n\Theta_n,~~~\mathcal M_n=\begin{bmatrix}
            z&Q_n\\
            R_n&z^{-1}
        \end{bmatrix},~~~\Theta_n=\begin{bmatrix}
            \theta_{1,n}\\
            \theta_{2,n}
        \end{bmatrix},\\
\label{sec2-time-evo}
 &    \Theta_{n,t}=\frac{1}{2i}\mathcal N_n\Theta_n,~~~\mathcal N_n=\begin{bmatrix}
            (z-z^{-1})^2-2Q_nR_{n-1}&2Q_nz-2z^{-1}Q_{n-1}\\
            2zR_{n-1}-2z^{-1}R_n&2R_nQ_{n-1}-(z-z^{-1})^2
        \end{bmatrix},
\end{align}
\end{subequations}
where \eqref{sec2-spectral} is known as the Ablowitz-Ladik (AL) spectral problem \cite{AL-JMP-1976,AL-JMP-1975},
which is a discretization of the  Ablowitz-Kaup-Newell-Segur (AKNS)
(or Zakharov-Shabat (ZS)-AKNS) spectral problem \cite{ZS-JETP-1972,ZS-JETP-1973,AKNS-PRL-1973}.
For the correspondence between the semi-discrete and continuous AKNS hierarchy,
one can refer to \cite{ZC-SAPM-2010a,ZC-SAPM-2010b}.

We call \eqref{sec2-AL2} the AL-2 system for short as it corresponds to the second-order AKNS equations \eqref{nls-c}.
It allows various reductions (also see \cite{AM-PRE-2014,DLZ-AMC-2018}):
\begin{subequations}\label{sec2-dNLS}
    \begin{eqnarray}
        i\partial_tQ_{n}=(1-\delta Q_nQ^*_n)(Q_{n+1}+Q_{n-1})-2Q_n,&& R_n=\delta Q_n^*,\label{sec2-dNLS-equ1}\\
        i\partial_tQ_{n}=(1-\delta Q_nQ^*_{-n})(Q_{n+1}+Q_{n-1})-2Q_n,&& R_n=\delta Q_{-n}^*,\label{sec2-dNLS-equ2}\\
        i\partial_tQ_{n}=(1-\delta Q_nQ_n(-t))(Q_{n+1}+Q_{n-1})-2Q_n,&& R_n=\delta Q_n(-t),\label{sec2-dNLS-equ3}\\
        i\partial_tQ_{n}=(1-\delta Q_nQ_{-n}(-t))(Q_{n+1}+Q_{n-1})-2Q_n,&& R_n=\delta Q_{-n}(-t),\label{sec2-dNLS-equ4}
    \end{eqnarray}
\end{subequations}
which are the classical, reverse-space, reverse-time and reverse-space-time sdNLS equations, respectively.
Here, $\delta=\pm1$, the function $Q$ with  reversed space, time and space-time are indicated by $Q_{-n}=Q(-n,t),~Q_{n}(-t)=Q(n,-t)$ and $Q_{-n}(-t)=Q(-n,-t)$, respectively.
As we have mentioned in the introduction section,
the classical sdNLS equation \eqref{sec2-dNLS-equ1}, i.e. \eqref{sdnls}
has been studied in great detail.
The nonlocal sdNLS equations, \eqref{sec2-dNLS-equ2}, \eqref{sec2-dNLS-equ3} and \eqref{sec2-dNLS-equ4}
were also solved via IST \cite{AM-PRE-2014,ALM-Nonl-2020},
Darboux transformation \cite{MZ-JMP-2016,XLZLL-AML-2017,XALX-SAPM-2024},
Hirota's bilinear method \cite{MZ-AML-2014,DLZ-AMC-2018}
and KP-reduction approach \cite{HC-CPL-2018,CFJ-WM-2020}.
Note that it is not easy if directly solving the nonlocal sdNLS equations,
and one may have to introduce trilinear equations rather than bilinear forms (see \cite{MZ-AML-2014} as an example).
We will see that the B-R approach does have advantages in solving nonlocal equations.

\section{Bilinear approach to the unreduced sdNLS} \label{sec-3}

In the B-R approach, the first step is to solve the unreduced sdNLS system, i.e. the AL-2 system \eqref{sec2-AL2},
by presenting its bilinear form and quasi double Casoratian solutions.

For the AL-2 system \eqref{sec2-AL2}, let $(q_n,r_n)$ be its any pair of solutions.
By introducing  transformation
\begin{equation}\label{sec3-tra1}
    Q_n=\frac{G_n}{F_n}, ~~~ R_n=\frac{H_n}{F_n},
\end{equation}
where $F_n, G_n$ and $H_n$ are all functions of $(n,t)$,
the AL-2 system \eqref{sec2-AL2} can be bilinearized into the following system
\begin{subequations}\label{sec3-bilinear-new}
    \begin{align}
    &F_n^2-(1-q_nr_n)F_{n+1}F_{n-1}=G_nH_n, \label{sec3-bilinear-new-equ1}\\
    &iD_tG_{n}\cdot F_n=(1-q_nr_n) (G_{n+1}F_{n-1}+G_{n-1}F_{n+1})-2G_nF_n, \label{sec3-bilinear-new-equ2}\\
    &iD_tF_n\cdot H_n=(1-q_nr_n) (F_{n+1}H_{n-1}+F_{n-1}H_{n+1})-2F_nH_n, \label{sec3-bilinear-new-equ3}
    \end{align}
    \end{subequations}
where $(q_n,r_n)$ is a given solution pair of the AL-2 system \eqref{sec2-AL2}
serving as nonzero backgrounds (see Remark \ref{rem-2}),
and $D$ is the Hirota bilinear operator  defined as \cite{H-PTP-1974}
\begin{equation*}
    D_t^s g(t)\cdot f(t)=(\partial_t-\partial_{t'})^s g(t)f(t')|_{t'=t}.
\end{equation*}
Note that when $q_n=r_n=0$, the above bilinear formula \eqref{sec3-bilinear-new} degenerates to the case of zero
background (see equation (8) in \cite{DLZ-AMC-2018}).

Solutions of the bilinear system \eqref{sec3-bilinear-new}
will be presented in terms of quasi double Casorati determinant (Casoratian).
Consider the following $(2m+2)$-th order vectors
\begin{equation}
    \Phi_n=(\phi_{1,n},\phi_{2,n},\cdots,\phi_{2m+2,n})^T, ~~~
    \Psi_n=(\psi_{1,n},\psi_{2,n},\cdots,\psi_{2m+2,n})^T,
\end{equation}
where $\phi_{j,n}$ and $\psi_{j,n}$ are functions of $(n,t)$.
Assume that $\Phi_n$ and $\Psi_n$ are defined by matrix equations
\begin{subequations}\label{sec3-phipsi}
    \begin{equation}
         \left(\begin{matrix}\Phi_{n+1}\\
            \Psi_{n+1}\end{matrix}\right)
         =M_n\left(\begin{matrix}\Phi_{n}\\
            \Psi_{n}\end{matrix}\right), ~~~
          2i\partial_t\left(\begin{matrix}\Phi_{n}\\
            \Psi_{n}\end{matrix}\right)
          = N_n \left(\begin{matrix}\Phi_{n}\\
            \Psi_{n}\end{matrix}\right),\label{sec3-phipsi-xt}
    \end{equation}
with
\begin{align}
        & M_n=\alpha_n^{-1/2}
        \left(\begin{matrix}
          A & q_nI_{2m+2}\\
          r_nI_{2m+2} & A^{-1}
           \end{matrix}\right), \label{sec3-mat-M}\\
        & N_n=\left(\begin{matrix}
         {(A-A^{-1})^2-(q_nr_{n-1}+r_nq_{n-1})I_{2m+2}} & 2Aq_n-2A^{-1}q_{n-1}\\
          2Ar_{n-1}-2A^{-1}r_{n} & (q_nr_{n-1}+r_nq_{n-1})I_{2m+2}-(A-A^{-1})^2
           \end{matrix}\right), \label{sec3-mat-N}
    \end{align}
\end{subequations}
where, $\alpha_n=1-q_n r_n$, $A\in \mathbb{C}_{(2m+2)\times (2m+2)}$,
$|A|\neq 0$, and $I_{k}$ is the $k$-th order identity matrix.
Define quasi double Casoratians composed by the above vectors $\Phi_n$ and $\Psi_n$:
\begin{subequations}\label{sec3-transform3}
    \begin{align}
    &    F(A,\Phi_n,\Psi_n)=\alpha_n^{(m+1)/2}|\Phi_{n+1},A^2\Phi_{n+1}, \cdots,A^{2m}\Phi_{n+1};
    \Psi_n,A^2\Psi_n, \cdots,A^{2m}\Psi_n|,\\
    &    G(A,\Phi_n,\Psi_n)= \alpha_n^{(m+1)/2}|\Phi_n,A\Phi_{n+1},A^3\Phi_{n+1}, \cdots,A^{2m+1}\Phi_{n+1};
    A \Psi_n,A^3\Psi_n, \cdots,A^{2m-1}\Psi_n|,\\
    &    H(A,\Phi_n,\Psi_n)= \alpha_n^{(m+1)/2}|A\Phi_{n+1},A^3\Phi_{n+1}, \cdots,A^{2m-1}\Phi_{n+1};
    \Psi_{n+1}, A\Psi_n,A^3\Psi_n, \cdots,A^{2m+1}\Psi_n|.
    \end{align}
\end{subequations}
If we use relations give by  the first equation in \eqref{sec3-phipsi-xt}, the above determinants
can alternatively be written as
\begin{subequations}\label{sec3-transform33}
    \begin{align}
    &    F(A,\Phi_n,\Psi_n)=|A\Phi_{n},A^3\Phi_{n}, \cdots,A^{2m+1}\Phi_{n};
    \Psi_n,A^2\Psi_n, \cdots,A^{2m}\Psi_n|, \label{3.7a}\\
    &    G(A,\Phi_n,\Psi_n)= |\Phi_n, A^2 \Phi_{n}, \cdots,A^{2m+2}\Phi_{n};
    A\Psi_n, A^3\Psi_n, \cdots,A^{2m-1}\Psi_n|+ (-1)^m q_n E(A,\Phi_n,\Psi_n),\\
    &    H(A,\Phi_n,\Psi_n)=|A^{2}\Phi_{n}, A^4 \Phi_{n}, \cdots, A^{2m}\Phi_{n};
    A^{-1}\Psi_{n}, A\Psi_n,  \cdots, A^{2m+1}\Psi_n|+ (-1)^m r_n E(A,\Phi_n,\Psi_n),
    \end{align}
where
\begin{equation}
E(A,\Phi_n,\Psi_n)=|\Phi_n, A^2 \Phi_{n},A^4\Phi_{n}, \cdots,A^{2m}\Phi_{n};
    A\Psi_n, A^3\Psi_n, \cdots,A^{2m+1}\Psi_n|.
\end{equation}
\end{subequations}

With the above notations, we come to the solutions of bilinear system \eqref{sec3-bilinear-new}.

\begin{theorem}\label{Th-1}
    The bilinear system \eqref{sec3-bilinear-new} has quasi double Casoratian solutions
\begin{equation}\label{sec3-transform2}
            F_n=F(A,\Phi_n,\Psi_n), ~~~
            G_n=G(A,\Phi_n,\Psi_n), ~~~
            H_n=H(A,\Phi_n,\Psi_n),
\end{equation}
where their entry vectors $\Phi_n$ and $\Psi_n$ satisfy matrix equations \eqref{sec3-phipsi}.
Note that matrix $A$  and any matrix that is similar to it   lead to same solutions
for $Q_n$ and $R_n$ through transformation \eqref{sec3-tra1}.
\end{theorem}

The proof is presented in Appendix \ref{app1-sec1}.

%%%%%%%%%%%%%%%%%%%%%%%%%%%%%%%%%%%%%%%%%%%

\section{Reductions and solutions}\label{sec-4}

The second step in the B-R approach is to impose constraints on  the vectors $\Phi_n$ and $\Psi_n$
so that the determinants $F_n, G_n$ and $H_n$ are constrained as well.
As a further result, $Q_n$ and $R_n$ defined by  $F_n, G_n$ and $H_n$
will satisfy some relations, which reduce the coupled AL-2 system to a single equation.
Such a reduction procedure is based on a technique developed in \cite{CZ-AML-2018,CDLZ-SAPM-2018,DLZ-AMC-2018},
which has proved effective in finding solutions for many equations, e.g.
\cite{SSZ-ND-2019,CLZ-AML-2019,LWZ-ROMP-2020,WWZ-2020,
W-ND-2021,LWZ-ROMP-2022,LWZ-SAPM-2022,WW-CNSNS-2022,WW-ND-2022,WWZ-CPB-2022,SLA-JPA-2023,
SWZ-AML-2021,ZFF-arX-2024}.

\subsection{Reduction technique}\label{sec-4-1}

In the following we take the case of the classical sdNLS equation \eqref{sec2-dNLS-equ1}
as an example to show how we get $R_n=\delta Q_n^*$
by imposing constraints on $\Phi_n$ and $\Psi_n$.

For $\Phi_n$ and $\Psi_n$ defined by \eqref{sec3-phipsi}, we impose a constraint
\begin{eqnarray}\label{sec3-phipsi-reduction1-case1}
    \Psi_n=T\Phi_n^*,
\end{eqnarray}
where $T\in\mathbb{C}_{(2m+2)\times(2m+2)}$ is a matrix to be fixed later.
Under the above constraint, together with  assumption $r_n=\delta q^*_n$ and
\begin{subequations}\label{sec3-at-relation-case1}
\begin{align}
  &      A^{-1}=TA^*T^{-1}, \label{sec3-at-relation-case1-A}\\
  &  TT^*=\delta I_{2m+2},  \label{sec3-at-relation-case1-T}
\end{align}
\end{subequations}
one can check that the matrix system \eqref{sec3-phipsi} can be reduced to the following:
\begin{subequations}\label{sec3-phipsi-reduction2-case1}
\begin{align}\label{sec3-phipsi-red2-spectral-case1}
& \Phi_{n+1}=\alpha_n^{-1/2}A\Phi_n+\alpha_n^{-1/2}q_nT\Phi^*_n,\\
&2i\partial_t\Phi_n=[(A-A^{-1})^2 -\delta (q_nq^*_{n-1}+q_{n-1}q^*_{n}) ]\Phi_n
+(2Aq_n-2A^{-1}q_{n-1})T\Phi^*_n.
\end{align}
\end{subequations}
where  $\alpha_n=1-\delta |q_n|^2$.
Note that if we introduce  $B\in\mathbb{C}_{(2m+2)\times(2m+2)}$ such that $A=e^B$,
the conditions in \eqref{sec3-at-relation-case1} are rewritten as
\begin{equation}\label{sec3-at-relation2-case1}
    -B=TB^*T^{-1}, ~~~  TT^*=\delta I_{2m+2}.
\end{equation}
With the constraints \eqref{sec3-phipsi-reduction1-case1} and \eqref{sec3-at-relation-case1},
we can rewrite the quasi double Casoratian $F_n$ in  \eqref{3.7a} as
\begin{align*}
%    F_n&= |A\Phi_{n},A^3\Phi_{n}, \cdots,A^{2m+1}\Phi_{n};\Psi_n,A^2\Psi_n \cdots,A^{2m}\Psi_n|  \\
 F_n   &= |A\Phi_n,A^3\Phi_n, \cdots,A^{2m+1}\Phi_n;T\Phi^*_n,
    T(A^{-2}\Phi_n)^*, \cdots,T(A^{-2m}\Phi_n)^*|,
\end{align*}
where we have made use of the relation $A^{-s}T=T(A^*)^s$ with $s\in \mathbb{Z}$  which follows from \eqref{sec3-at-relation-case1-A}.
Then, taking its complex conjugation and using \eqref{sec3-at-relation-case1-T}, one obtains
\begin{align}
        F_n^*&= |(A\Phi_n)^*,(A^3\Phi_n)^*,\cdots,(A^{2m+1}\Phi_n)^*;
        T^*\Phi_n,T^*A^{-2}\Phi_n,\cdots,T^*A^{-2m}\Phi_n| \nonumber\\
        &=\delta^{m+1}|A^{2m+1}T|^{-1}|T(A^{-2m}\Phi_n)^*,T(A^{-2m+2}\Phi_n)^*,\cdots,T\Phi_n^*;
        A^{2m+1}\Phi_n,A^{2m-1}\Phi_n,\cdots,A\Phi_n| \nonumber\\
        &= (-\delta)^{m+1}|A^{2m+1}T|^{-1}F_n. \label{F-F}
\end{align}
Similarly, we can derive
\begin{equation}
H_n^*=-(-\delta)^m|A^{2m+1}T|^{-1}G_n.
\label{G-H}
\end{equation}
Thus, from \eqref{sec3-tra1} we arrive at
\begin{eqnarray}
        R_n^*=\frac{H_n^*}{F_n^*}=\frac{-(-\delta)^m|A^{2m+1}T|^{-1}G_n}{(-\delta)^{m+1}|A^{2m+1}T|^{-1}F_n}
        =\delta \frac{G_n}{F_n}=\delta Q_n,
\end{eqnarray}
which is the reduction to get the classical sdNLS equation \eqref{sec2-dNLS-equ1}.

To summarize, for a given solution $q_n$ of equation \eqref{sec2-dNLS-equ1}
and $A$ and $T$ satisfying \eqref{sec3-at-relation-case1},
once we get $\Phi_n$ from \eqref{sec3-phipsi-reduction2-case1}, get $\Psi_n$ from \eqref{sec3-phipsi-reduction1-case1},
and use them to define $F_n$ and $G_n$ as in \eqref{sec3-transform33},
then, $Q_n=G_n/F_n$ provides a solution for the classical sdNLS equation \eqref{sec2-dNLS-equ1}.
In addition, we have a remark on \eqref{sec3-phipsi-reduction1-case1} and \eqref{sec3-at-relation-case1}.

\begin{remark}\label{rem-1}
    The condition \eqref{sec3-at-relation-case1} indicates $|A||A^*|=|T||T^*|=1$,
    which means both $|A|$ and $|T|$ are the points on the unit circle of the complex plane.
    In addition, it is easy to verified that condition \eqref{sec3-at-relation-case1}
    is equivalent to (a same form)
    \begin{equation}
    A^{-1}=\hat TA^*\hat T^{-1}, ~~~ \hat T\hat T^*=\delta  I_{2m+2},
    \label{R1}
    \end{equation}
    where
    \begin{equation}
    \hat T=e^{-i\gamma }A^{-\mu}T
    \label{R2}
    \end{equation}
    with $\gamma\in \mathbb R$ and $\mu\in \mathbb Z$.
    Meanwhile, under the transformation \eqref{R2}, the constraint \eqref{sec3-phipsi-reduction1-case1}
    is mapped to
    \begin{eqnarray}
        e^{i\gamma } A^\mu\Psi_n=\hat T\Phi_n,
    \end{eqnarray}
    where $A$ and $\hat T$ obey \eqref{R1}.
    A special case is to choose $\gamma=0$, $\mu=1-2m$ and  $q_n=r_n=0$,
    which brings us those quasi double Casoratians $F_n$ and $G_n$ presented in \cite{DLZ-AMC-2018}.
    Similar discussions can be extended to the nonlocal cases.
\end{remark}

Next, we come to the nonlocal case.
To achieve the nonlocal reduction in \eqref{sec2-dNLS-equ2}, i.e. $R_n=\delta Q_{-n}^*$,
we start from an assumption $r_n=\delta q_{-n}^*$,
impose constraint
\begin{eqnarray}\label{sec3-phipsi-reduction1-case2}
    \Psi_n=T\Phi_{1-n}^*
\end{eqnarray}
and require  $A$ and $T$ to satisfy
\begin{eqnarray}\label{sec3-at-relation-case2}
        A=TA^*T^{-1},~~~ TT^*=-\delta I_{2m+2}.
\end{eqnarray}
It can be checked that   the matrix system \eqref{sec3-phipsi}  reduces to
\begin{subequations}\label{sec3-phipsi-reduction2-case2}
\begin{align}\label{sec3-phipsi-red2-spectral-case2}
& \Phi_{n+1}=\alpha_n^{-1/2}A\Phi_n+\alpha_n^{-1/2}q_nT\Phi^*_{1-n},\\
& 2i\partial_t\Phi_{n}=[(A-A^{-1})^2-\delta (q_nq_{1-n}^*+q_{n-1}q_{-n}^*)]\Phi_n+(2Aq_n-2A^{-1}q_{n-1})T\Phi_{1-n}^*,
\end{align}
\end{subequations}
with $\alpha_n=1-\delta q_nq_{-n}^*$.
In this nonlocal case, $F_n$ is presented as
\[F_n=\alpha_n^{(m+1)/2}|\Phi_{n+1},A^2\Phi_{n+1}, \cdots, A^{2m}\Phi_{n+1};
T\Phi^*_{1-n},A^2T\Phi^*_{1-n}, \cdots, A^{2m}T\Phi^*_{1-n}|.
\]
Using \eqref{sec3-at-relation-case2}   it is easy to obtain
\begin{align*}
 F_{-n}^*&=(\alpha_{-n}^{(m+1)/2})^*
                   |\Phi_{1-n}^*,(A^*)^2\Phi_{1-n}^*, \cdots, (A^*)^{2m}\Phi_{1-n}^*;
                     T^*\Phi_{n+1}, (A^*)^2T^*\Phi_{n+1}, \cdots, (A^*)^{2m}T^*\Phi_{n+1}|\\
    &=\alpha_{n}^{(m+1)/2} (-\delta)^{m+1} |T|^{-1}
                   |T\Phi_{1-n}^*,A^2T\Phi_{1-n}^*, \cdots, A^{2m}T\Phi_{1-n}^*;
                     \Phi_{n+1}, A^2\Phi_{n+1}, \cdots, A^{2m}\Phi_{n+1}|\\
    &=\alpha_{n}^{(m+1)/2} (-\delta)^{m+1} |T|^{-1}
                   |T\Phi_{1-n}^*,A^2T\Phi_{1-n}^*, \cdots, A^{2m}T\Phi_{1-n}^*;
                     \Phi_{n+1}, A^2\Phi_{n+1}, \cdots, A^{2m}\Phi_{n+1}|\\
    &=\delta^{m+1}|T|^{-1}F_n.
\end{align*}
Similarly, we have
\[ H_n=\alpha_n^{(m+1)/2}|A\Phi_{n+1},A^3\Phi_{n+1},\cdots,A^{2m-1}\Phi_{n+1};
T\Phi^*_{-n}, AT\Phi^*_{1-n},\cdots,A^{2m+1}T\Phi^*_{1-n}|
\]
and
\[
        H_{-n}^*=\delta^{m}|T|^{-1}G_n.
\]
It then follows that
\[
R_{-n}^*=\frac{H_{-n}^*}{F_{-n}^*}=\frac{\delta^{m}|T|^{-1}G_n}{\delta^{m+1}|T|^{-1}F_n}
=\delta G_n/F_n=\delta Q_n,
\]
which gives rise to the nonlocal reduction for  the reverse-space sdNLS equation \eqref{sec2-dNLS-equ2}.

We can continue to investigate constraints and reductions for the other two equations in \eqref{sec2-dNLS}.
In the following we skip the details and just list main results
in Table \ref{tab-1} for the four equations in \eqref{sec2-dNLS}.
Note that in the Table $\mathbf{0}$ and $\mathbf{I}$ respectively
denote zero matrix and identity matrix of $2m+2$ order.

\begin{table}[!htbp]
    \centering
    \captionsetup{font={small}}
    \caption{Constraints and reductions for \eqref{sec2-dNLS}}\label{tab-1}
    \begin{tabular}{|c|c|c|c|c|}
    \hline
         eq.      &            $(q_n,r_n)$           &  constraint            &  $F_n, G_n, H_n$    & $\Phi_n$    \\
    \hline
   \multirow{3}{*}{\eqref{sec2-dNLS-equ1}} &  \multirow{3}{*}{$r_n=\delta q_n^*$ }
   & $\Psi_n=T\Phi_n^*$ & $F_n^*=(-\delta)^{m+1}|A^{2m+1}T|^{-1}F_n$ &
    \multirow{3}{*}{\eqref{sec3-phipsi-reduction2-case1}}\\
  &  &  $A^{-1}T-TA^*=\mathbf 0,~ TT^*=\delta \mathbf{I}$ & $ H_n^*=-(-\delta)^m|A^{2m+1}T|^{-1}G_n$ & \\
   &  & $BT+TB^*=\mathbf{0}, ~ TT^*=\delta \mathbf{I}$ &   &\\
  \hline
   \multirow{3}{*}{\eqref{sec2-dNLS-equ2}} &  \multirow{3}{*}{$r_n=\delta q_{-n}^*$ }
   & $ \Psi_n=T\Phi_{1-n}^*$ & $F_{-n}^*=\delta^{m+1}|T|^{-1}F_n $ &
   \multirow{3}{*}{\eqref{sec3-phipsi-reduction2-case2}}\\
  &  &  $ AT-TA^*=\mathbf{0},~ TT^*=-\delta \mathbf{I}$ & $ H_{-n}^*=\delta^{m}|T|^{-1}G_n $ &\\
   &  & $ BT-TB^*=\mathbf{0}, ~ TT^*=-\delta \mathbf{I}$ &   &\\
  \hline
   \multirow{3}{*}{\eqref{sec2-dNLS-equ3}} &  \multirow{3}{*}{$r_n=\delta q_{n}(-t)$}
   & $\Psi_n=T\Phi_{n}(-t) $ & $F_{n}(-t)= (-\delta)^{m+1}|A^{2m+1}T|^{-1}F_n(t) $ &
   \multirow{3}{*}{\eqref{sec3-phipsi-reduction2-case3}}\\
  &  &  $ A^{-1}T-TA=\mathbf0,~ TT=\delta \mathbf{I} $ & $ H_{n}(-t)=-(-\delta)^{m}|A^{2m+1}T|^{-1}G_n(t) $ &\\
   &  & $BT+TB=\mathbf{0},~ TT=\delta \mathbf{I}$ &   &\\
  \hline
   \multirow{3}{*}{\eqref{sec2-dNLS-equ4}} &  \multirow{3}{*}{$r_n=\delta q_{-n}(-t)$ }
   & $  \Psi_n=T\Phi_{1-n}(-t)$ & $F_{-n}(-t)=\delta^{m+1}|T|^{-1}F_n(t) $ &
   \multirow{3}{*}{\eqref{sec3-phipsi-reduction2-case4}}\\
  &  &  $ AT-TA=\mathbf{0},~ TT=-\delta \mathbf{I} $ & $ H_{-n}(-t)= \delta^{m}|T|^{-1}G_n(t) $ &\\
   &  & $BT-TB=\mathbf{0},~ TT=-\delta \mathbf{I} $ &   &\\
  \hline
\end{tabular}
\end{table}

Here in Table \ref{tab-1}, for \eqref{sec2-dNLS-equ3} and \eqref{sec2-dNLS-equ4},
vector $\Phi_n$ is determined respectively by
\begin{subequations}\label{sec3-phipsi-reduction2-case3}
\begin{align}\label{sec3-phipsi-red2-spectral2}
& \Phi_{n+1}=\alpha_n^{-1/2}A\Phi_n+\alpha_n^{-1/2}q_nT\Phi_{n}(-t),
~~~~~ (\alpha_n=1-\delta q_nq_n(-t)),\\
& 2i\partial_t\Phi_{n}=[(A-A^{-1})^2-\delta (q_nq_{n-1}(-t)+q_{n-1}q_{n}(-t)) ]\Phi_n
+(2Aq_n-2A^{-1}q_{n-1})T\Phi_{n}(-t),
\end{align}
\end{subequations}
and
\begin{subequations}\label{sec3-phipsi-reduction2-case4}
\begin{align}\label{sec3-phipsi-red2-spectral2}
&\Phi_{n+1}=\alpha_n^{-1/2}A\Phi_n+\alpha_n^{-1/2}q_nT\Phi_{1-n}(-t),
~~~~~ (\alpha_n=1-\delta q_nq_{-n}(-t)), \\
&2i\partial_t\Phi_{n}=[(A-A^{-1})^2-\delta (q_nq_{1-n}(-t)+q_{n-1}q_{-n}(-t)) ]\Phi_n
+(2Aq_n-2A^{-1}q_{n-1})T\Phi_{1-n}(-t).
\end{align}
\end{subequations}

Let us summarize this subsection.

\begin{theorem}\label{Th-2}
Solutions of the four sdNLS equations in \eqref{sec2-dNLS} are given by
\begin{equation}\label{Q-FG}
Q_n=\frac{G_n}{F_n},
\end{equation}
where $F_n$ and $G_n$ are quasi double Casoratians defined in \eqref{sec3-transform3} (or \eqref{sec3-transform33})
and $\Phi_n$ and $\Psi_n$ are given as in Table \ref{tab-1}.
\end{theorem}

\begin{remark}\label{rem-2}
For the classical sdNLS equation, using relations \eqref{F-F} and \eqref{G-H},
the unreduced bilinear system reduces to
\begin{subequations}\label{sec3-bilinear-sdnls}
\begin{align}
&F_n F^*_n - (1-\delta |q_n|^2) F_{n+1}F_{n-1}^*=\delta G_n G_n^*, \label{sec3-bilinear-sdnls1}\\
&iD_tG_{n}\cdot F_n=(1-\delta |q_n|^2) (G_{n+1}F_{n-1}+G_{n-1}F_{n+1})-2G_nF_n.
\end{align}
\end{subequations}
This (with $\delta =-1$) is different from the known one
(see the formula at the bottom of page 17 in \cite{OY-JPA-2014})
even when assuming $F_n=F_n^*$.
When $\delta=-1$, from  \eqref{Q-FG} and \eqref{sec3-bilinear-sdnls1} we find
\begin{equation}\label{QQ}
|Q_n|^2=(1+|q_n|^2)\frac{F_{n+1}F_{n-1}^*}{F_n F_n^*}-1.
\end{equation}
In general, $F_n$ and $F_{n+1}$ go to same result when $n\to \pm \infty$
(except some special cases where $F_n$  is asymptotically dominated by
periodic functions of $n$, e.g. the Ahkmediev breather, see Sec.\ref{sec-5-1-1}).
Thus, in most of cases we have
\begin{equation*}
|Q_n|\sim |q_n|,~~ ~ (n\to \pm \infty).
\end{equation*}
In this sense, we say $(q_n, r_n)$ are background solutions of the AL-2 system \eqref{sec2-AL2}.
\end{remark}

\subsection{Explicit forms of matrices $B$ and $T$}\label{sec-4-2}

The reductions now boil down to solving matrix equations in Table \ref{tab-1}.
In this subsection we look for explicit forms of $B$ and $T$.
The equations for $B$ and $T$ in Table \ref{tab-1} can be  unified to be
the following two types (cf.\cite{CDLZ-SAPM-2018}):
\begin{equation}\label{sec3-con-at1}
    BT+\sigma TB^*=\mathbf{0}, ~~   TT^*=\sigma\delta \mathbf{I}, ~~~ \sigma,\delta=\pm 1,
\end{equation}
and
\begin{equation}\label{sec3-con-at2}
    BT+\sigma TB=\mathbf{0},~~  TT=\sigma\delta \mathbf{I}, ~~~ \sigma,\delta=\pm 1.
\end{equation}
We consider the following special $2\times 2$ block matrix forms:
\begin{eqnarray}\label{sec4-bt-sol}
    B=\left(\begin{matrix}
        K_1 & \mathbf{0}_{m+1}\\
        \mathbf{0}_{m+1} & K_4
    \end{matrix}\right),~~
     T=\left(\begin{matrix}
        T_1 & T_2\\
        T_3 & T_4
    \end{matrix}\right)
\end{eqnarray}
where $T_j, K_j \in \mathbb{C}_{(m+1)\times (m+1)}$ matrices,
and $\mathbf{0}_{m+1}$ stands for the zero matrix of $m+1$ order.
In this case, solutions to equations \eqref{sec3-con-at1} and \eqref{sec3-con-at2}
 are given in Table \ref{tab-2} and Table \ref{tab-3}, respectively.

\begin{table}[htbp]
    \centering
    \captionsetup{font={small}}
    \caption{$T$ and~$B$ for equation~\eqref{sec3-con-at1}}\label{tab-2}
    \begin{tabular}{|c|c|c|c|c|}
    \hline
             eq.                     &            $(\sigma,\delta)$           &  $T$              &  $B$                                  \\
    \hline
    \multirow{4}{*}{\eqref{sec3-con-at1}} & $(1,1)$  &
    $T_{1}=T_{4}=\mathbf{0}_{m+1},T_{2}=T_{3}= {I}_{m+1}$  &
    \multirow{2}{*}{$K_{1}=\mathbf{K}_{m+1}, K_{4}=-\mathbf{K}^*_{m+1}$}    \\
        \cline{2-3}
                              & $(1,-1)$ & $T_{1}=T_{4}=\mathbf{0}_{m+1},T_{2}=-T_{3}={I}_{m+1}$  & \\
        \cline{2-4}
  & $(-1,1)$ &
    $T_{1}=T_{4}=\mathbf{0}_{m+1},T_{2}=-T_{3}={I}_{m+1}$ &
    \multirow{2}{*}{$K_{1}=\mathbf{K}_{m+1}, K_{4}=\mathbf{K}^*_{m+1}$} \\
        \cline{2-3}
                                     &  $(-1,-1)$ &  $T_{1}=T_{4}=\mathbf{0}_{m+1},T_{2}=T_{3}={I}_{m+1}$  &  \\
        \cline{1-4}
    \hline
    \end{tabular}
    \end{table}

    \begin{table}[htbp]
    \centering
    \captionsetup{font={small}}
    \caption{$T$ and~$B$ for equation~\eqref{sec3-con-at2}}\label{tab-3}
    \begin{tabular}{|c|c|c|c|c|}
    \hline
      eq.                            &            $(\sigma,\delta)$           &  $T$              &  $B$                                  \\
    \hline
    \multirow{4}{*}{\eqref{sec3-con-at2} } & $(1,1)$  &
    $T_{1}=T_{4}=\mathbf{0}_{m+1},T_{2}=T_{3}= {I}_{m+1}$  &
    \multirow{2}{*}{$K_{1}=\mathbf{K}_{m+1}, K_{4}=-\mathbf{K}_{m+1}$}    \\
        \cline{2-3}
                                     & $(1,-1)$ & $T_{1}=T_{4}=\mathbf{0}_{m+1},T_{2}=-T_{3}= {I}_{m+1}$  & \\
        \cline{2-4}
      & $(-1,1)$ &
    $T_{1}=-T_{4}=i {I}_{m+1},T_{2}=T_{3}=\mathbf{0}_{m+1}$ &
    \multirow{2}{*}{$K_{1}=\mathbf{K}_{m+1}, K_{4}=\mathbf{H}_{m+1}$} \\
        \cline{2-3}
                                     &  $(-1,-1)$ & $T_{1}=-T_{4}= {I}_{m+1},T_{2}=T_{3}=\mathbf{0}_{m+1}$  &  \\
        \cline{1-4}
    \hline
    \end{tabular}
    \end{table}

Here in Table \ref{tab-2}, $\mathbf{K}_{m+1}\in \mathbb{C}_{(m+1)\times (m+1)}$.
In addition,  equation \eqref{sec3-con-at1} with $(\sigma,\delta)=(-1,1)$
admits a real solution in the form \eqref{sec4-bt-sol} where
\begin{subequations}\label{sec4-bt-sol-spe1}
    \begin{eqnarray}
        &&K_1=\mathbf{K}_{m+1},~~K_4=\mathbf{H}_{m+1}, ~~~
        \mathbf{K}_{m+1},\mathbf{H}_{m+1}\in \mathbb{R}_{(m+1)\times(m+1)},\\
        &&T_1=\pm T_4=I_{m+1}, T_2=T_3=\mathbf{0}_{m+1}.
    \end{eqnarray}
\end{subequations}

Since matrix $B$ and any matrix which is  similar to it
lead to same $Q_n$ and $R_n$,
in practice, we only need to consider the canonical form of matrix $B$, which is composed by
\begin{eqnarray}
        \mathbf{K}_{m+1}=\mathrm{Diag}(J_{h_1}(k_1),J_{h_2}(k_2),\cdots,J_{h_s}(k_s))
\end{eqnarray}
with $\sum_{i=1}^sh_i=m+1$,
where $J_{h}(k)$ is a Jordan block defined by
\begin{eqnarray}\label{Jordan}
        J_{h}(k)=\left(\begin{matrix}
        k & 0 & 0 & \ldots & 0 & 0\\
        1 & k & 0 & \ldots & 0 & 0\\
        \ldots & \ldots & \ldots & \ldots & \ldots & \ldots\\
        0 & 0 & 0 & \ldots & 1 & k
        \end{matrix}\right)_{h\times h}.
\end{eqnarray}
The two elementary cases of the canonical form are
\begin{eqnarray}\label{K-diag}
    \mathbf{K}_{m+1}=\mathrm{Diag}(k_1,k_2,\cdots,k_{m+1})
\end{eqnarray}
and
\begin{eqnarray}\label{K-jord}
    \mathbf{K}_{m+1}=J_{m+1}(k_1).
\end{eqnarray}

\subsection{Explicit expressions of $\Phi_n$ and $\Psi_n$}\label{sec-4-3}

\subsubsection{The unreduced case 1: plane wave background}\label{sec-4-3-1}

To construct $\Phi_n$ and $\Psi_n$,
we first consider the following system
\begin{subequations}\label{sec4-phipsi}
\begin{align}
& \left(\begin{matrix}\phi_{n+1}\\
            \psi_{n+1}\end{matrix}\right)
            =\alpha_n^{-1/2}
            \left(\begin{matrix}e^k&q_n\\r_n&e^{-k}
            \end{matrix}\right)
            \left(\begin{matrix}
            \phi_{n}\\
            \psi_{n}\end{matrix}\right),
            ~~~~ \alpha_n=1-q_n r_n,\label{sec4-phipsi-spectral}\\
& 2i\left(\begin{matrix}\phi_{n}\\
            \psi_{n}\end{matrix}\right)_t
            =\left(\begin{matrix}{(e^k-e^{-k})^2-q_nr_{n-1}-q_{n-1}r_n}&2e^kq_n-2e^{-k}q_{n-1}\\
                2e^kr_{n-1}-2e^{-k}r_{n}&q_nr_{n-1}+q_{n-1}r_n-(e^k-e^{-k})^2
            \end{matrix}\right)
            \left(\begin{matrix}\phi_{n}\\
            \psi_{n}\end{matrix}\right),~~~\label{sec4-phipsi-time}
\end{align}
\end{subequations}
where $q_n, r_n$ are given solutions of the AL-2 system \eqref{sec2-dNLS},
$\phi_n$ and $\psi_n$ are scalar functions.
Consider the following plane wave solutions of \eqref{sec2-dNLS}:
\begin{eqnarray}\label{qr-pw}
    q_n=a_0e^{2i \delta a_0^2 t},&r_n=\delta a_0 e^{-2i\delta a_0^2 t}, & \delta=\pm 1,
\end{eqnarray}
where $a_0$ is a real constant.
In this case,  equation set \eqref{sec4-phipsi} admits a solution pair
\begin{subequations}\label{phipsi}
\begin{eqnarray}
    &&\phi_n(k,c,d)=(c\,e^{\lambda n+\eta t}+d\,e^{-(\lambda n+\eta t)})e^{i\delta a^2_0 t},\\
    &&\psi_n(k,c,d)=(-c\,\xi(k)e^{\lambda n+\eta t}+d\,\xi(-k) e^{-(\lambda n+\eta t)})e^{-i\delta a^2_0 t},
\end{eqnarray}
\end{subequations}
with
\begin{subequations}\label{4.29}
\begin{eqnarray}
    &&e^{\lambda}=e^{\lambda(k)}=\frac{e^k+e^{-k}+\sqrt{(e^k-e^{-k})^2+4\delta a_0^2}}
    {2\sqrt{1-\delta a_0^2}}, \label{4.29a}\\
    &&\eta=\eta(k)=-\frac i2(e^k-e^{-k})\sqrt{(e^k-e^{-k})^2+4\delta a_0^2}, \label{4.29b}\\
    &&\xi=\xi(k)=\frac{e^k-e^{-k}-\sqrt{(e^k-e^{-k})^2+4\delta a_0^2}}{2a_0},
\end{eqnarray}
\end{subequations}
and $c, d$ being constants (or functions of $k$).
Using the above $\phi_n(k,c,d)$ and $\psi_n(k,c,d)$ we define vectors
\begin{subequations}\label{hat-check-phipsi1}
\begin{align}
&\Phi_n=(\phi_n(k_1,c_1,d_1),\phi_n(k_2,c_2,d_2),\cdots,\phi_n(k_{2m+2},c_{2m+2},d_{2m+2}))^T,\\
&\Psi_n=(\psi_n(k_1,c_1,d_1),\psi_n(k_2,c_2,d_2),\cdots,\psi_n(k_{2m+2},c_{2m+2},d_{2m+2}))^T.
\end{align}
\end{subequations}
Then,  $\Phi_n$ and $\Psi_n$ provide solutions for the matrix equations \eqref{sec3-phipsi}
where $(q_n, r_n)$ are given in \eqref{qr-pw}
and $A$ is
\begin{equation}\label{B-diag}
A=e^B,~~ B=\mathrm{Diag}(k_1,k_2,\cdots, k_{2m+2}).
\end{equation}
If we take plane wave $(q_n,r_n)$ as in \eqref{qr-pw} and
\begin{equation}\label{B-jord}
A=e^B,~~ B=J_{2m+2}(k),
\end{equation}
then \eqref{sec3-phipsi} admits solutions
\begin{subequations}\label{hat-check-phipsi2}
\begin{align}
&\Phi_n=\left(\phi_n(k,c,d),\frac{\partial_{k}}{1!}\phi_n(k,c,d),\cdots,
\frac{\partial^{2m+1}_{k}}{(2m+1)!}\phi_n(k,c,d)\right)^T,\\
&\Psi_n=\left(\psi_n(k,c,d),\frac{\partial_{k}}{1!}\psi_n(k,c,d), \cdots,
\frac{\partial^{2m+1}_{k}}{(2m+1)!}\psi_n(k,c,d)\right)^T,
\end{align}
\end{subequations}
where $\phi_n(k,c,d)$ and $\psi_n(k,c,d)$ are defined in \eqref{phipsi}.

\subsubsection{The unreduced case 2: hyperbolic function background}\label{sec-4-3-2}

Other than the plane wave solution \eqref{qr-pw},
the AL-2 system \eqref{sec2-dNLS} has a second simple solution (cf.\cite{XALX-SAPM-2024})
\begin{equation}\label{qr-tanh}
    q_n=a_0\tanh(\mu n+\nu )e^{2 i a_0^2 t},
    ~~~ r_n= a_0 \tanh(\mu n+\nu)e^{-2 i a_0^2 t},
\end{equation}
where $\mu\in \mathbb{R}$, $\nu\in \mathbb{C}$  and  $a_0=\tanh(\mu)$.
The equation set \eqref{sec4-phipsi} with the above $(q_n,r_n)$ allows the following solutions
\begin{subequations}\label{4.30}
\begin{equation}
\phi_n(k,c,d)=\hat\gamma_n\hat\phi_n(k,c,d),~~~
\psi_n(k,c,d)=\hat\gamma_n\hat\psi_n(k,c,d),
\end{equation}
where
\begin{eqnarray}\label{4.38b}
    \hat\gamma_{n}=\prod^{n-1}_{s=-\infty}\sqrt{\frac{1-a_0^2}{\alpha_s}},
    ~~~ \alpha_s=1-q_sr_s,
\end{eqnarray}
\begin{align}
        \hat\phi_{n}(k,c,d)=\,&c[\xi(-k)e^k+\tanh(\mu n+\nu-\mu)]e^{\lambda n+\eta t+ia_0^2t}\nonumber\\
        &+d[-\xi(k)e^k+\tanh(\mu n+\nu-\mu)]e^{-\lambda n-\eta t+i a_0^2t},\label{4.38c}\\
        \hat\psi_{n}(k,c,d) =\,&c[-\xi(k)-e^k\tanh(\mu n+\nu-\mu)]e^{\lambda n+\eta t-ia_0^2t}\nonumber\\
         &+d[\xi(-k)-e^k\tanh(\mu n+\nu-\mu)]e^{-\lambda n-\eta t-i a_0^2t}, \label{4.38d}
\end{align}
and
\begin{eqnarray}
    &&e^{\lambda}=e^{\lambda(k)}=\frac{e^k+e^{-k}+\sqrt{(e^k-e^{-k})^2+4a_0^2}}{2\sqrt{1-a_0^2}},\\
    &&\eta=\eta(k)=-\frac{i}{2}(e^k-e^{-k})\sqrt{(e^k-e^{-k})^2+4a_0^2},\\
    &&\xi(k)=\frac{e^k-e^{-k}-\sqrt{(e^k-e^{-k})^2+4a_0^2}}{2a_0}.
\end{eqnarray}
\end{subequations}
Here, again,  $c$ and $d$ are constants (or functions of $k$).

With $\phi_n(k,c,d)$ and $\psi_n(k,c,d)$ defined in \eqref{4.30},
the vectors $\Phi_n$ and $\Psi_n$ in the form \eqref{hat-check-phipsi1} provide solutions
for \eqref{sec3-phipsi} with $A$ given in \eqref{B-diag} and  $(q_n,r_n)$ given in \eqref{qr-tanh},
while the vectors $\Phi_n$ and $\Psi_n$ in the form \eqref{hat-check-phipsi2} provide solutions
for \eqref{sec3-phipsi} with $A$ given in \eqref{B-jord}.

\subsubsection{The reduced cases}\label{sec-4-3-3}

For the reduced equations in \eqref{sec2-dNLS}, the vectors $\Phi_n$ and $\Psi_n$
for their solutions can be determined by considering the constraints in Table \ref{tab-1}
and their solutions given in Table \ref{tab-2} and Table \ref{tab-3}.
For convenience, we present $\Phi_n$ and $\Psi_n$ in the following form:

\begin{subequations}\label{sec4-phipsi-relation}
\begin{equation}
\Phi_n=\left(\begin{array}{c}
                  \Phi_n^+\\
                  \Phi_n^-
                  \end{array}\right),~~~
\Psi_n=\left(\begin{array}{c}
                  \Psi_n^+\\
                  \Psi_n^-
                  \end{array}\right),
\end{equation}
where
\begin{eqnarray}
 \Phi_n^\pm=(\Phi_{1,n}^\pm,\cdots,\Phi_{m+1,n}^\pm)^T, ~~~
 \Psi_n^\pm=(\Psi_{1,n}^\pm,...,\Psi_{m+1,n}^\pm)^T,
\end{eqnarray}
\end{subequations}
with $\Phi_{j,n}^\pm$ and $\Psi_{j,n}^\pm$ being scalar functions.

For equation \eqref{sec2-dNLS-equ1}, \eqref{sec2-dNLS-equ2} and \eqref{sec2-dNLS-equ3},
their corresponding matrix $T$ is block skew diagonal,
it is possible to express $\Phi_{j,n}^\pm$ and $\Psi_{j,n}^\pm$ through
$\Phi_{j,n}^+$ and $\Psi_{j,n}^+$.
When $\mathbf{K}_{m+1}$ is diagonal as given in \eqref{K-diag}, we let
\begin{eqnarray}
    \Phi_{j,n}^+=\phi_n(k_j,c_j,d_j),~~~  \Psi_{j,n}^+=\psi_n(k_j,c_j,d_j), ~~~ j=1,2,\cdots,m+1,
\end{eqnarray}
where $\phi_n(k,c,d)$ and $\psi_n(k,c,d)$ can be either \eqref{phipsi} or \eqref{4.30},
depending on $(q_n,r_n)$.
When $\mathbf{K}_{m+1}=J_{m+1}(k)$ as given in \eqref{K-jord}, we let
\begin{eqnarray}
    \Phi_{j,n}^+=\frac{\partial^{j-1}_k \phi(k,c,d)}{(j-1)!},~~~
    \Psi_{j,n}^+=\frac{\partial^{j-1}_k \psi(k,c,d)}{(j-1)!}, ~~~ j=1,2,\cdots,m+1.
\end{eqnarray}
We list out $\Phi_n$ and $\Psi_n$ in Table \ref{tab-4} for more explicity.

\begin{table}[htbp]
    \centering
    \captionsetup{font={small}}
    \caption{$\Phi_n$ and $\Psi_n$  for \eqref{sec2-dNLS-equ1}, \eqref{sec2-dNLS-equ2}
    and \eqref{sec2-dNLS-equ3}}\label{tab-4}
    \begin{tabular}{|c|c|}
    \hline
             eq.                     &            $\Phi_n$ and $\Psi_n$       \\
    \hline
   \eqref{sec2-dNLS-equ1} &
    $\Phi_n=\left(\begin{array}{c}
                  \Phi_n^+\\
                  \Psi_n^{+*}
                  \end{array}\right), ~~
   \Psi_n=\left(\begin{array}{c}
                  \Psi_n^+\\
                  \delta\,\Phi_n^{+*}
                  \end{array}\right)$   \\
   \hline
   \eqref{sec2-dNLS-equ2} &
   $\Phi_n=\left(\begin{array}{c}
                  \Phi_n^+\\
                  \Psi_{1-n}^{+*}
                  \end{array}\right), ~~
   \Psi_n=\left(\begin{array}{c}
                  \Psi_n^+\\
                  -\delta\,\Phi_{1-n}^{+*}
                  \end{array}\right)$  \\
     \hline
   \eqref{sec2-dNLS-equ3} &
   $\Phi_n=\left(\begin{array}{c}
                  \Phi_n^+\\
                 \Psi_n^{+}(-t)
                  \end{array}\right), ~~
   \Psi_n=\left(\begin{array}{c}
                  \Psi_n^+\\
                  \delta\,\Phi_n^{+}(-t)
                  \end{array}\right)$   \\
     \hline
    \end{tabular}
    \end{table}

For equation \eqref{sec2-dNLS-equ4} and the special reduction \eqref{sec4-bt-sol-spe1},
their corresponding matrix $T$ is block diagonal.
This means we can  express $\Psi_{j,n}^\pm$ through $\Phi_{j,n}^\pm$.
When $\mathbf{K}_{m+1}$ and $\mathbf{H}_{m+1}$ are diagonal as given in \eqref{K-diag}, we take
\begin{eqnarray}
    \Phi_{j,n}^+=\phi_n(k_j,c^+_j,d^+_j),~~~  \Phi_{j,n}^-=\phi_n(h_j,c^-_j,d^-_j), ~~~ j=1,2,\cdots,m+1,
\end{eqnarray}
where $\phi_n(k,c,d)$ and $\psi_n(k,c,d)$ are given by \eqref{phipsi}.
When $\mathbf{K}_{m+1}=J_{m+1}(k)$ and  $\mathbf{H}_{m+1}=J_{m+1}(h)$ as in \eqref{K-jord}, we take
\begin{eqnarray}
    \Phi_{j,n}^+=\frac{\partial^{j-1}_k \phi(k,c^+,d^+)}{(j-1)!},~~~
    \Phi_{j,n}^-=\frac{\partial^{j-1}_h \psi(h,c^-,d^-)}{(j-1)!}, ~~~ j=1,2,\cdots,m+1.
\end{eqnarray}
$\Phi_n$ and $\Psi_n$ of this case are listed in Table \ref{tab-5}.

\begin{table}[htbp]
    \centering
    \captionsetup{font={small}}
    \caption{$\Phi_n$ and $\Psi_n$  for equation \eqref{sec2-dNLS-equ4} and the special reduction \eqref{sec4-bt-sol-spe1}}\label{tab-5}
    \begin{tabular}{|c|c|c|}
    \hline
             eq.                     &            $\Phi_n$ and $\Psi_n$  & $d_j^+$ and $d_j^-$      \\
    \hline
    \multirow{2}{*}{\eqref{sec2-dNLS-equ4}} &
    \multirow{2}{*}{
    $\Phi_n=\left(\begin{array}{c}
                  \Phi_n^+\\
                  \Phi_n^{-}
                  \end{array}\right), ~~
   \Psi_n=\left(\begin{array}{c}
                  i^{(\delta+1)/2}\Phi^+_{1-n}(-t)\\
                  -i^{(\delta+1)/2}\Phi^-_{1-n}(-t)
                  \end{array}\right)$ }
    & $d_j^+=i^{(\delta+1)/2}c^+_j\xi(k_j)e^{\lambda(k_j)}$\\
    & & $d_j^-=-i^{(\delta+1)/2}c^-_j\xi(h_j)e^{\lambda(h_j)}$  \\
   \hline
   \multirow{2}{*}{\eqref{sec4-bt-sol-spe1}}&
   \multirow{2}{*}{
    $\Phi_n=\left(\begin{array}{c}
                  \Phi_n^+\\
                  \Phi_n^{-}
                  \end{array}\right), ~~
   \Psi_n=\left(\begin{array}{c}
                  \Phi^{+*}_{1-n}\\
                  \pm\Phi^{-*}_{1-n}
                  \end{array}\right)$ }
        & $d_j^+=-c^{+*}_j\xi(k_j)e^{\lambda(k_j)}$\\

    &&$d_j^-=-c^{-*}_j\xi(h_j)e^{\lambda(h_j)}$\\
     \hline
    \end{tabular}
    \end{table}

Here in Table \ref{tab-5}, $\xi(k)$ and $e^{\lambda}$ are defined as in \eqref{4.29}.

%%%%%%%%%%%%%%%%%%%%%%%%%%%%%%%

\section{Dynamics of solutions}\label{sec-5}

In this section, we are going to analyze and illustrate some solutions for the classical sdNLS equation
and the reverse-space sdNLS equation.

\subsection{The classical focusing dNLS   with a plane wave background}\label{sec-5-1}

For the focusing sdNLS equation \eqref{sec2-dNLS-equ1} (with $\delta=-1$),
the squared envelop $|Q_n|^2$, which has been given in \eqref{QQ}, can also be written as
\begin{eqnarray}\label{5.1}
    |Q_n|^2=(1+|q_n|^2)\frac{  F_{n+1}  F_{n-1}}{  F_n^2}-1,
\end{eqnarray}
where
\begin{eqnarray}\label{5.2}
      F_n=  |A\Phi_n,A^3\Phi_n,\cdots,A^{2m+1}\Phi_n;
    T(\Phi_n)^*,T(A^{-2}\Phi_n)^*,\cdots,T(A^{-2m}\Phi_n)^*|,
\end{eqnarray}
and $\Phi_n$ should be taken accordingly from Sec.\ref{sec-4-3-3}.
As we have explained in Remark \ref{rem-2}, $|q_n|$ can be viewed as a background of $|Q_n|$
and we call $q_n$ is a background solution of the focusing sdNLS equation \eqref{sec2-dNLS-equ1}.
The analysis in this subsection is for the plane wave background in \eqref{qr-pw}, i.e.
\begin{equation}\label{q-pw}
 q_n=a_0e^{-2i  a_0^2 t}.
\end{equation}

\subsubsection{Breathers}\label{sec-5-1-1}

We will see that with the plane wave background \eqref{q-pw},
there is no usual solitons for the focusing sdNLS equation,
instead, the typical solutions behave like breathers.

\vskip 5pt
\noindent
\textbf{Case 1: $\mathbf{K}_{m+1}$ being a diagonal matrix}

\begin{figure}[ht!]
\centering
\subfigure[ ]{\label{Fig-1a}
\begin{minipage}[t]{0.40\linewidth}
\centering
\includegraphics[width=2.3in]{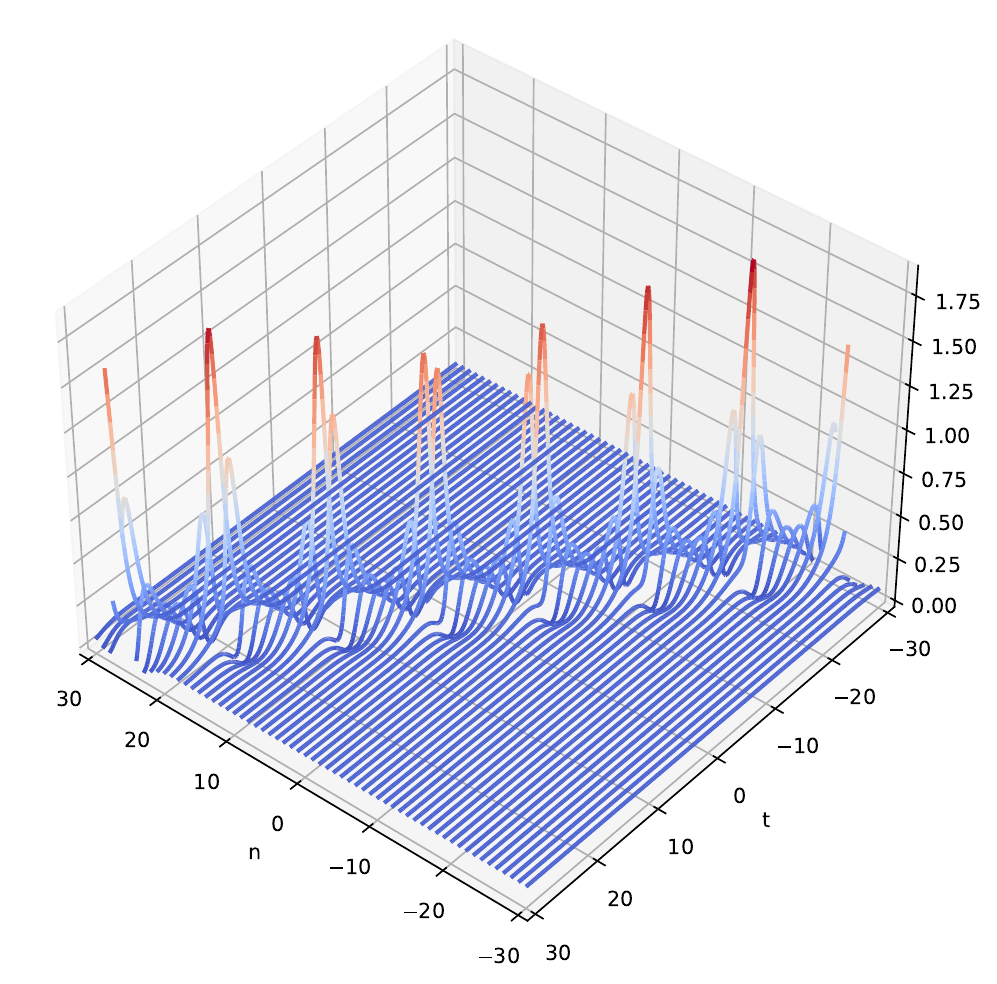}
\end{minipage}%
}%
\subfigure[ ]{\label{Fig-1b}
\begin{minipage}[t]{0.40\linewidth}
\centering
\includegraphics[width=2.5in]{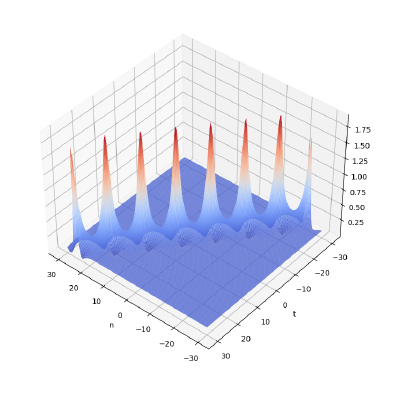}
\end{minipage}%
}%

\subfigure[ ]{\label{Fig-1c}
\begin{minipage}[t]{0.40\linewidth}
\centering
\includegraphics[width=2.5in]{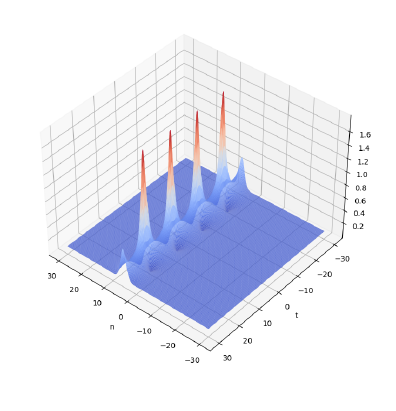}
\end{minipage}%
}%
\subfigure[ ]{\label{Fig-1d}
\begin{minipage}[t]{0.40\linewidth}
\centering
\includegraphics[width=2.5in]{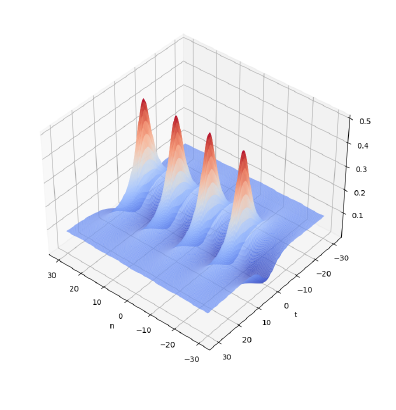}
\end{minipage}%
}%
\caption{\label{Fig-1}
Shape and motion of the squared envelop of  one breather solution of the focusing sdNLS equation.
(a) and (b) a moving breather   for ${k_1}=\ln(1.5-0.2i),~c_1=d_1=1,~a_0=0.3$.
(c) a Kuznetsov-Ma breather   for ${k_1}=\ln(1.5),~c_1=d_1=1,~a_0=0.3$.
(d) an Ahkmediev breather   for ${k_1}=\ln(1.2),~c_1=d_1=1,~a_0=0.3$.
}
\end{figure}

When $m=0$,  we have  $\mathbf{K}_{1}=k_1$ and
\begin{eqnarray}
    \Phi_n=(\phi_n(k_1,c_1,d_1),
    (\psi_n(k_1,c_1,d_1))^*)^T,
\end{eqnarray}
where $\phi_n$ and $\psi_n$ are defined as in \eqref{phipsi} with $\delta=-1$.
This yields
\begin{align*}
    e^{k_1^*} F_n&=-|e^{k}\phi_n(k_1,c_1,d_1)|^2-|\psi_n(k_1,c_1,d_1)|^2  \\
    & = -C_1 e^{2(a_1n+a_2t)}-C_2 e^{-2(a_1n+a_2t)}-C_3 e^{2i(b_1n+b_2t)}- C_4 e^{-2i(b_1n+b_2t)},
\end{align*}
where
%\begin{subequations}
\begin{align*}
& C_1=|c_1|^2 (|e^{k_1}|^2+|\xi({k_1})|^2),   ~~~
 C_2=|d_1|^2 (|e^{k_1}|^2+|\xi(-{k_1})|^2),\\
& C_3=c_1d_1^*(|e^{k_1}|^2-\xi({k_1})\xi(-{k_1}^*)),  ~~~
C_4 =c^*_1d_1(|e^{k_1}|^2-\xi({k_1}^*)\xi(-{k_1})),\\
&  \lambda=a_1+ib_1, ~~~ \eta=a_2+ib_2,~~~~ a_j,b_j\in\mathbb{ R}, ~(j=1,2).
\end{align*}
%\end{subequations}
Note that  $C_1,C_2\in \mathbb{R}$, $C_3=C_4^*$,  and $\lambda, \xi(k)$ and $\eta$ are formulated in \eqref{4.29}.
Rewrite the above $ F_n$ as
\begin{equation}\label{Fn}
e^{k_1^*} F_n=D_1 \cosh (2(a_1n+a_2t))+D_2 \sinh (2(a_1n+a_2t))
+D_3 \cos (2(b_1n+b_2t)) +D_4 \sin (2(b_1n+b_2t)),
\end{equation}
where
\[D_1=-2(C_1+C_2),~~ D_2=-2(C_1-C_2),~~ D_3=-(C_3+C_4), ~~D_4=i(C_3-C_4).\]
This indicates that
$|Q_n|^2$ is dominated by a `solitary' wave traveling parallel to the line $a_1n+a_2t=0$,
coupled by a oscillating behavior due to the trigonometric functions.
Note that $D_3^2+D_4^2\neq 0$ unless in trivial solutions.
Such a combination gives rise to a breather, as depicted in Fig.\ref{Fig-1a} and \ref{Fig-1b}.\footnote{
In principle, we should provide all figures as Fig.\ref{Fig-1a} for $n\in \mathbb{Z}$.
We use figures with smooth surfaces (e.g. Fig.\ref{Fig-1b}) for a better look.
}
To see more details, we take a close look at $\lambda$ and $\eta$ defined in \eqref{4.29}, i.e.
\begin{align*}
& e^{\lambda}=e^{a_1+ib_1}
=\frac{e^{k_1}+e^{-k_1}+\sqrt{(e^{k_1}-e^{-k_1})^2-4  a_0^2}}
    {2\sqrt{1+ a_0^2}},  \\
&  \eta=a_2+ib_2=-\frac i2(e^{k_1}-e^{-k_1})\sqrt{(e^{k_1}-e^{-k_1})^2-4  a_0^2}.
\end{align*}
One special case is to take $k_1$ such that
\begin{eqnarray}
    {k_1} \in \mathbb R, && (e^{k_1}-e^{-{k_1}})^2>4a_0^2,
\end{eqnarray}
which yields   $a_2=0$ and a stationary breather perpendicular to $n$-axis, which is called
a Kuznetsov-Ma breather, cf.\cite{K-SPD-1977,M-SAPM-1979}, as shown in Fig.\ref{Fig-1c}.
Another special case is from
\begin{eqnarray}
    {k_1} \in \mathbb R, && (e^{k_1}-e^{-{k_1}})^2<4a_0^2,
\end{eqnarray}
or
\begin{eqnarray}
    {k_1} \in i\mathbb R.
\end{eqnarray}
It then follows that $|e^{\lambda}|=1$, which indicates  $a_1=0$ and
leads to a breather perpendicular to $t$-axis, known as an  Ahkmediev breather,
cf.\cite{AEK-TMP-1987}, as  shown in Fig.\ref{Fig-1d}.

\captionsetup[figure]{labelfont={bf},name={Fig.},labelsep=period}

\begin{figure}[ht!]
\centering
\subfigure[ ]{
\begin{minipage}[t]{0.32\linewidth}
\centering
\includegraphics[width=2.0in]{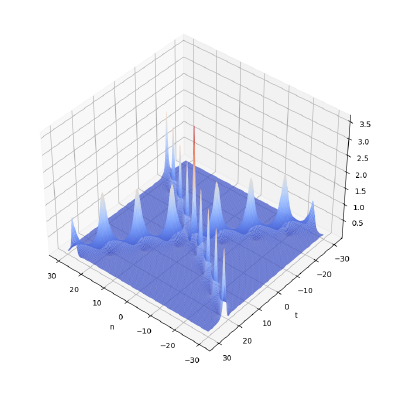}
\end{minipage}%
}%
\subfigure[ ]{
\begin{minipage}[t]{0.32\linewidth}
\centering
\includegraphics[width=2.0in]{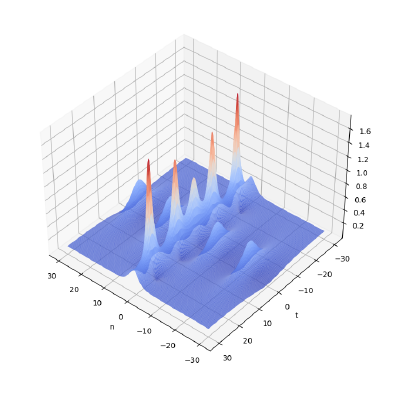}
\end{minipage}%
}%
\subfigure[ ]{
\begin{minipage}[t]{0.32\linewidth}
\centering
\includegraphics[width=2.0in]{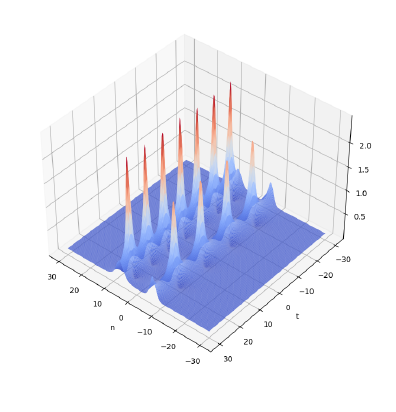}
\end{minipage}%
}%
\caption{\label{Fig-2}
Shape and motion of the squared envelop of  two-breather solution of the focusing sdNLS equation.
(a) interaction of two traveling breathers
for $k_1=\ln(1.5-0.2i),k2=\ln(1.5+0.4i),~c_1=c_2=d_1=d_2=1,~a_0=0.3$.
(b) interaction of a Kuznetsov-Ma breather and an Ahkmediev breather for $k_1=\ln(1.5),k_2=\ln(1.2),~c_1=c_2=d_1=d_2=1,~a_0=0.3$.
(c) two parallel Kuznetsov-Ma breathers for $k_1=\ln(1.5),~k_2=\ln(1.6),~c_1=d_2=10,~c_2=d_1=1,~a_0=0.3$.
}
\end{figure}

Two-breather solutions can be obtained by taking $m=1$, which results in
the following component vector
\begin{eqnarray}
    \Phi_n=(\phi_n(k_1,c_1,d_1),\phi_n(k_{2},c_{2},d_{2}),
    (\psi_n(k_1,c_1,d_1))^*, (\psi_n(k_{2},c_{2},d_{2}))^*)^T.
\end{eqnarray}
The squared envelop $|Q_n|^2$ is defined via \eqref{5.1} and \eqref{5.2}.
Based on the analysis of one-breather solutions,
it is expected various types of two-breather interactions.
For example, interaction of two traveling breathers, interaction of a Kuznetsov-Ma breather and an Ahkmediev breather,
and interaction of two Kuznetsov-Ma breathers.
They are all illustrated in Fig.\ref{Fig-2}.

\captionsetup[figure]{labelfont={bf},name={Fig.},labelsep=period}

\vskip 5pt
\noindent
\textbf{Case 2: $\mathbf{K}_{m+1}$ being a Jordan matrix}

\begin{figure}[ht!]
\centering
\subfigure[ ]{
\begin{minipage}[t]{0.32\linewidth}
\centering
\includegraphics[width=2.0in]{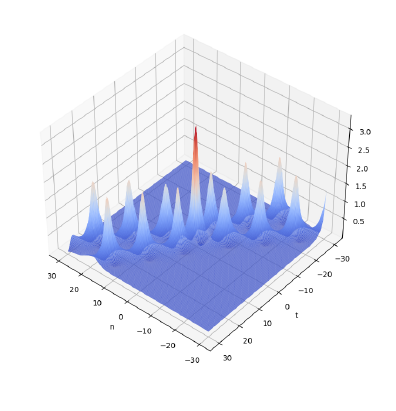}
\end{minipage}%
}%
\subfigure[ ]{
\begin{minipage}[t]{0.32\linewidth}
\centering
\includegraphics[width=2.0in]{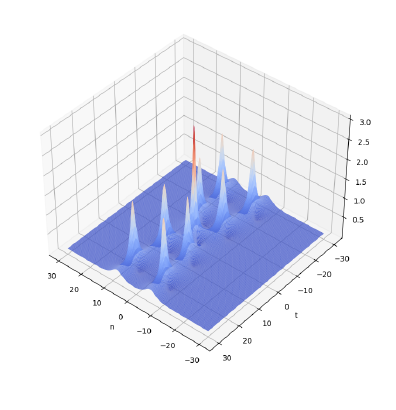}
\end{minipage}%
}%
\subfigure[ ]{
    \begin{minipage}[t]{0.32\linewidth}
    \centering
    \includegraphics[width=2.0in]{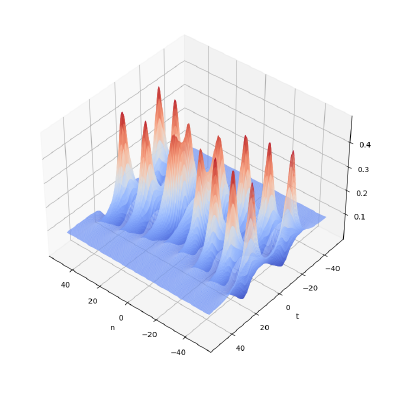}
    \end{minipage}%
    }%
\caption{\label{Fig-3} Shape and motion of the squared envelop of (Jordan matrix)
breather solution of the focusing sdNLS equation.
(a) traveling breathers for $k_1=\ln(1.5-0.2i),~c_1=d_1=1,~a_0=0.3$.
(b) Kuznetsov-Ma breathers for $k_1=\ln(1.5),~c_1=d_1=1,~a_0=0.3$.
(b) Akhmediev breathers for $k_1=\ln(1.2),~c_1=d_1=1,~a_0=0.3$.
}
\end{figure}

For $m=1$, $\Phi_n$ can be written as the following,
\begin{eqnarray}
    \Phi_n=(\phi_n(k_1,c_1,d_1),\partial_{k_1}\phi_n(k_{1},c_{1},d_{1}),
    \psi_n(k_1,c_1,d_1)^*,(\partial_{k_1}\psi_n(k_{1},c_{1},d_{1}))^*)^T.
\end{eqnarray}
The squared envelop $|Q_n|^2$ is defined via \eqref{5.1} and \eqref{5.2}.
Fig.\ref{Fig-3} shows traveling breathers,
 Kuznetsov-Ma breathers and Akhmediev breathers obtained by using Jordan matrix $J_2(k_1)$,  respectively.

\captionsetup[figure]{labelfont={bf},name={Fig.},labelsep=period}

\subsubsection{Rogue waves}\label{sec-5-1-2}

To achieve rogue waves, we introduce a parameter $\kappa$ by
\begin{equation}\label{kappa}
\kappa={\sqrt{(e^k-e^{-k})^2-4a_0^2}}.
\end{equation}
For the functions $\phi_n$ and $\psi_n$ defined in \eqref{phipsi},
in terms of $\kappa$ we have the following expressions for the involved elements:
\begin{subequations}
\begin{align}
& e^k=\frac{1}{2}\sqrt{\kappa^2+4+4a_0^2}+\frac12\sqrt{\kappa^2+4a^2_0},~~~
e^{-k}= \frac{1}{2}\sqrt{\kappa^2+4+4a_0^2}-\frac12\sqrt{\kappa^2+4a^2_0},\\
&  e^{\lambda}=\frac{\kappa+\sqrt{\kappa^2+4+4a_0^2}}{\sqrt{4+4a_0^2}}, ~~~ e^{-\lambda}=\frac{-\kappa+\sqrt{\kappa^2+4+4a_0^2}}{\sqrt{4+4a_0^2}},\\
&    \eta=-\frac i2\kappa\sqrt{\kappa+4a_0^2},\\
&    \xi(k)=\tilde{\xi}(\kappa)=\frac{1}{2a_0}\Bigl(\sqrt{\kappa^2+4a_0^2}-\kappa\Bigr),~~~
\xi(-k)=-\tilde{\xi}(-\kappa).
\end{align}
\end{subequations}
We also assume $c$ and $d$ are functions of $\kappa$ and satisfy
\begin{equation}\label{cd-kappa}
c(\kappa)=-d(-\kappa).
\end{equation}
With these settings, $\phi_n$ and $\psi_n$  in \eqref{phipsi} can be written as
\begin{subequations}\label{phipsi-kappa}
\begin{align}
&\phi_n =\Biggl[ c(\kappa)
\Biggl(\frac{\kappa+\sqrt{\kappa^2+4+4a_0^2}}{\sqrt{4+4a_0^2}}\Biggr)^n e^{\eta t}
-c(-\kappa) \Biggl(\frac{\kappa+\sqrt{\kappa^2+4+4a_0^2}}{\sqrt{4+4a_0^2}}\Biggr)^{-n} e^{-\eta t}
\Biggr] e^{- i a^2_0  t},\\
&\psi_n =\Biggl[-c(\kappa) \tilde{\xi}(\kappa) \Biggl(\frac{\kappa+\sqrt{\kappa^2+4+4a_0^2}}{\sqrt{4+4a_0^2}}\Biggr)^n e^{\eta t}
+c(-\kappa)\tilde{\xi}(-\kappa)
\Biggl(\frac{\kappa+\sqrt{\kappa^2+4+4a_0^2}}{\sqrt{4+4a_0^2}}\Biggr)^{-n} e^{-\eta t}
\Biggr] e^{ i a^2_0 t}.
\end{align}
\end{subequations}
To meet \eqref{cd-kappa}, we assume   $c(\kappa)$ and $d(\kappa)$
to be the following series:
\begin{eqnarray}
    c(\kappa)=\sum_{j=0}^\infty s_j\kappa^j,~~ d(\kappa)=\sum_{j=0}^\infty (-1)^{j+1}s_j\kappa^j,
\end{eqnarray}
where $\{s_j\}$ can be arbitrary complex numbers.
Both $\phi_n$ and $\psi_n$ in \eqref{phipsi-kappa} are odd functions of $\kappa$,
which can be expanded as
\begin{eqnarray}
\phi_n=\sum_{j=0}^\infty R_{2j+1}\kappa^{2j+1}, && \psi_n=\sum_{j=0}^\infty S_{2j+1}\kappa^{2j+1},
\end{eqnarray}
where
\begin{equation}
R_{2j+1}=\frac{\partial_{\kappa}^{2j+1}}{(2j+1)!}\phi_n|_{\kappa=0}, ~~
S_{2j+1}=\frac{\partial_{\kappa}^{2j+1}}{(2j+1)!}\psi_n|_{\kappa=0}.
\end{equation}
Define
\begin{eqnarray}\label{5.20}
    \Phi_n=(R_1,R_3, \cdots, R_{2m+1}, S_1^*, S_3^*, \cdots, S_{2m+1}^*)^T
\end{eqnarray}
and denote
\begin{eqnarray}\label{5.21}
    e^{\mathbf{K}_{m+1}}=\left(\begin{matrix}
        \zeta_0&0&0&\cdots&0\\
        \zeta_2&\zeta_0&0&\cdots&0\\
        \vdots&\vdots&\vdots&\ddots&\vdots\\
        \zeta_{2m}&\zeta_{2m-2}&\zeta_{2m-4}&\cdots&\zeta_0
    \end{matrix}\right),
\end{eqnarray}
where $\{\zeta_{2j}\}$ are defined by
\begin{eqnarray}\label{5.22}
    e^k=\sum_{j=0}^\infty\zeta_{2j}\kappa^{2j}.
\end{eqnarray}
Then, it can be proved that $\Phi_n$ defined in \eqref{5.20} satisfies the equation set \eqref{sec3-phipsi-reduction2-case1}
where $\delta=-1$,
\begin{equation}\label{A-rw}
 A=\mathrm{Diag}(e^{\mathbf{K}_{m+1}},e^{-\mathbf{K}_{m+1}^*}),
\end{equation}
$q_n$ is given by \eqref{q-pw} and
$T=\left(\begin{matrix}
        \mathbf{0}_{m+1} & I_{m+1}\\
        -I_{m+1} & \mathbf{0}_{m+1}
    \end{matrix}\right)$.

Thus, we achieve explicit rational solutions (rogue waves) for the focusing sdNLS equation:
\begin{subequations}\label{rw-solu}
\begin{equation}
Q_n=\frac{G_n}{F_n},
\end{equation}
where $F_n$ and $G_n$ are the quasi double Casoratians composed by the above
$\Phi_n, A$ and $T$:
\begin{align}
    F_n=& |A\Phi_n,A^3\Phi_n,\cdots,A^{2m+1}\Phi_n;
    T\Phi_n^*,A^{2}T\Phi_n^*,\cdots,A^{2m}T\Phi_n^*|, \label{Fn-RW}\\
    G_n=& |\Phi_n, A^2 \Phi_{n}, \cdots,A^{2m+2}\Phi_{n};
     AT\Phi_n^*, A^{3}T\Phi_n^*,\cdots, A^{2m-1}T \Phi_n^*| \nonumber\\
    &+ (-1)^m q_n  |\Phi_n, A^2 \Phi_{n}, \cdots,A^{2m}\Phi_{n};
    AT\Phi_n^*, A^{3}T\Phi_n^*,\cdots, A^{2m+1}T \Phi_n^*|.
\end{align}
\end{subequations}
The squared envelop $|Q_n|^2$ is given by the formula \eqref{5.1} with $F_n$ in \eqref{Fn-RW} and
$q_n$ in \eqref{q-pw}.

\begin{figure}[ht!]
\centering
\subfigure[ ]{\label{Fig-4a}
\begin{minipage}[t]{0.40\linewidth}
\centering
\includegraphics[width=2.5in]{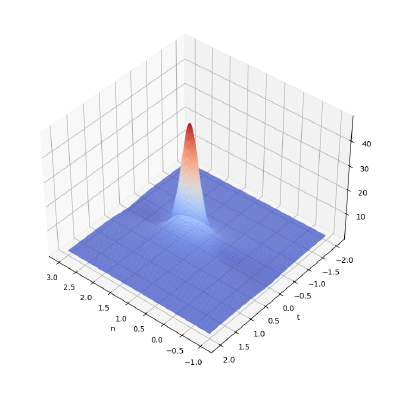}
\end{minipage}%
}%
\subfigure[ ]{\label{Fig-4b}
\begin{minipage}[t]{0.40\linewidth}
\centering
\includegraphics[width=2.5in]{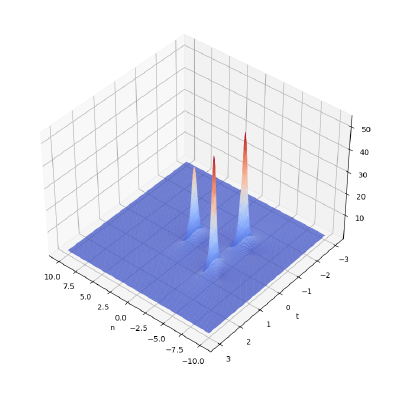}
\end{minipage}%
}%
\caption{ Shape and motion of the squared envelop of  rational solutions of the focusing sdNLS equation.
(a) the first order rogue wave for $a_0=c(\kappa)=1$.
(b) the second order rogue wave for $c(\kappa)=1+\kappa$ and $a_0=1$.
}\label{Fig-4}
\end{figure}

When $m=0$, we get the simplest rogue wave solution
\begin{subequations}
\begin{eqnarray}
Q_n=-a_0\left(1+\frac{8 it-\frac{8i}{a_0}\Im(\frac{s_1}{s_0})-\frac{2}{a_0^2}}{U_n}\right) e^{-2i a^2_0 t},
\end{eqnarray}
with
\begin{align}
    U_n=& 8\Bigl(a_0t-\Im\Bigl(\frac{s_1}{s_0}\Bigr)\Bigr)^2
    +\biggl(\frac{n}{\sqrt{1+a_0^2}}+2\Re\Bigl(\frac{s_1}{s_0}\Bigr)\biggr)^2
    +\biggl(\frac{n}{\sqrt{1+a_0^2}}-\frac{1}{a_0}+2\Re\Bigl(\frac{s_1}{s_0}\Bigr)\biggr)^2 \nonumber\\
    &+\frac{1}{\sqrt{1+a_0^2}}\biggl(\frac{2n}{\sqrt{1+a_0^2}}-\frac{1}{a_0}
    +4\Re\Bigl(\frac{s_1}{s_0}\Bigr)\biggr),
\end{align}
\end{subequations}
and its squared envelop is
\begin{eqnarray}
    |Q_n|^2=\frac{a_0^2[(U_n-\frac{2}{a_0^2})^2+64(t-\frac{1}{a_0}\Im(\frac{s_1}{s_0}))^2]}{U_n^2},
\end{eqnarray}
which is depicted in Fig.\ref{Fig-4a}.
Here and after, we denote $z=\Re(z)+i\Im(z)$ for $z\in \mathbb{C}$.

The explicit formula \eqref{rw-solu} allows us to calculate high order solution easily.
We just illustrate a second order ($m=1$) rogue wave in  Fig.\ref{Fig-4b}
while skip its expression.

\captionsetup[figure]{labelfont={bf},name={Fig.},labelsep=period}

%\newpage

%%%%%%%%%%%%%%%%%%%%%%%%%%%%%%%%%%%%

\subsection{The reverse-space defocusing sdNLS  with a plane wave background}\label{sec-5-2}

\subsubsection{Solitons and doubly periodic solutions}\label{sec-5-2-1}

For the reverse-space defocusing sdNLS equation \eqref{sec2-dNLS-equ2} with a plane wave background
$q_n$ give in \eqref{qr-pw} where $\delta=1$,
its one-soliton solution comes from the case $m=0$ and $\mathbf K_1=k_1$.
In this case, we have
\begin{eqnarray}
    \Phi_n=(\phi_n(k_1,c_1,d_1),(\psi_{1-n}(k_1,c_1,d_1))^*)^T
\end{eqnarray}
where $\phi_n$ and $\psi_n$ are defined in \eqref{phipsi}.
Then we have
%\begin{subequations}
\begin{align*}
\alpha_n^{-1/2}F_n & =|\Phi_{n+1};\Psi_n|  \\
    &=C_1e^{2(a_2t + i b_1 n)}+C_2e^{-2(a_2t + i b_1 n)}+C_3e^{2(a_1n+ib_2t)}+C_4e^{-2(a_1n+ib_2t)},\\
\alpha_n^{-1/2}G_n & =|\Phi_{n};A\Phi_{n+1}|  \\
  &  =\left( D_1e^{2(a_2t + i b_1 n)}+D_2e^{-2(a_2t + i b_1 n)}
  +D_3e^{2(a_1n+i b_2t)}+D_4e^{-2(a_1n+i b_2t)}\right)
  e^{2i a_0^2 t},
\end{align*}
%\end{subequations}
where $\alpha_n=1-q_nq^*_{-n}$,
with
%\begin{subequations}
\begin{align*}
    & C_1=-|c_1|^2(|e^\lambda|^2+|\xi(k_1)|^2),   ~~~
     C_2=-|d_1|^2(|e^{-\lambda}|^2+|\xi(-k_1)|^2) ,\\
    & C_3=-c_1d_1^*(e^{\lambda-\lambda^*}-\xi(k_1)(\xi(-k_1))^*),  ~~~
    C_4 =-c^*_1d_1(e^{\lambda^*-\lambda}-(\xi(k_1))^*\xi(-k_1)),\\
    &D_1=-|c_1|^2(\xi(k_1))^*(e^{k_1^*}-e^{k_1}e^{\lambda+\lambda^*}),~~~
    D_2=|d_1|^2(\xi(-k_1))^*(e^{k_1^*}-e^{k_1}e^{-(\lambda+\lambda^*)}),\\
    &D_3=c_1d_1^*(\xi(-k_1))^*(e^{k_1^*}-e^{k_1}e^{\lambda-\lambda^*}), ~~~
    D_4=-c_1^*d_1(\xi(k_1))^*(e^{k_1^*}-e^{k_1}e^{-(\lambda-\lambda^*)}),\\
    &  \lambda=\lambda(k_1)=a_1+ib_1, ~~~ \eta=\eta(k_1)=a_2+ib_2,~~~~ a_j,b_j\in\mathbb{ R}, ~(j=1,2),
\end{align*}
%\end{subequations}
and $\lambda, \eta$ and $\xi$ are defined in \eqref{4.29}.
One-soliton solution is then given by
\begin{equation}\label{1ss}
Q_n=\frac{G_n}{F_n}.
\end{equation}

\begin{figure}[ht!]
\centering
\subfigure[ ]{
\begin{minipage}[t]{0.32\linewidth}
\centering
\includegraphics[width=2.0in]{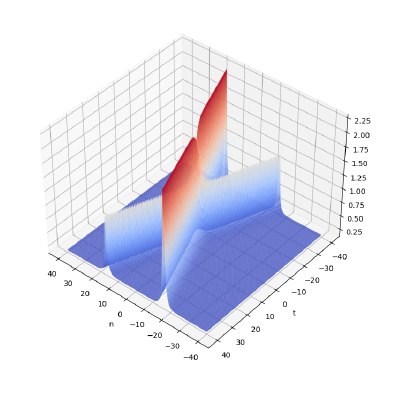}
\end{minipage}\label{Fig-5a}
}%
\subfigure[ ]{
\begin{minipage}[t]{0.32\linewidth}
\centering
\includegraphics[width=2.0in]{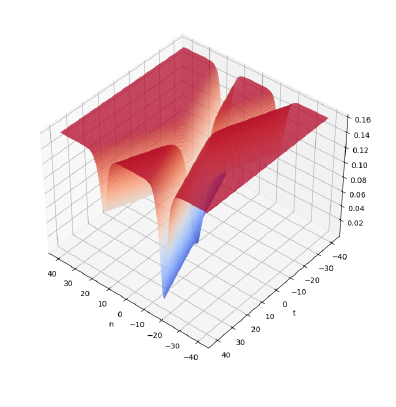}
\end{minipage}\label{nonlocal-plane-figs-2}
}%
\subfigure[ ]{
\begin{minipage}[t]{0.32\linewidth}
\centering
\includegraphics[width=2.0in]{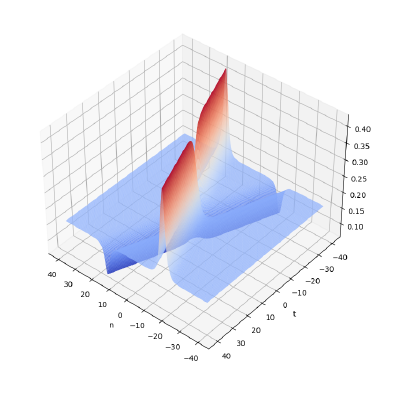}
\end{minipage}\label{nonlocal-plane-figs-3}
}%

\subfigure[ ]{
\begin{minipage}[t]{0.32\linewidth}
\centering
\includegraphics[width=2.0in]{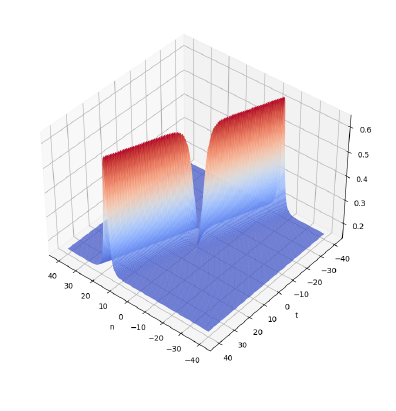}
\end{minipage}\label{nonlocal-plane-figs-4}
}%
\subfigure[ ]{
\begin{minipage}[t]{0.32\linewidth}
\centering
\includegraphics[width=2.0in]{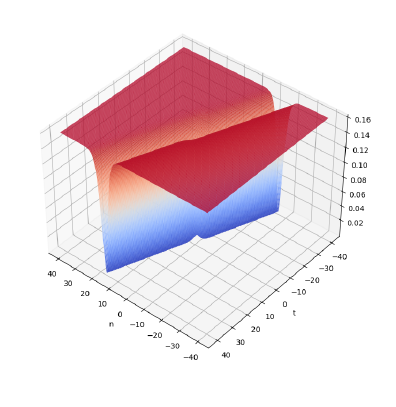}
\end{minipage}\label{Fig-5e}
}%
\subfigure[ ]{
\begin{minipage}[t]{0.32\linewidth}
\centering
\includegraphics[width=2.0in]{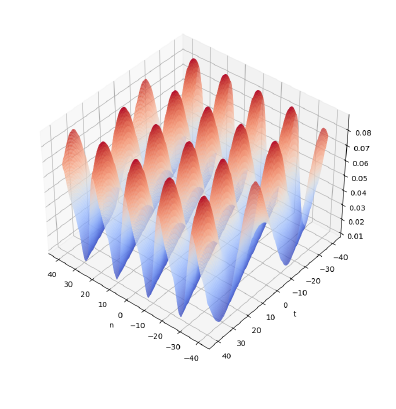}
\end{minipage}\label{Fig-5f}
}%
\caption{\label{Fig-5}
Shape and motion of the squared envelope of the solutions of the reverse-space defocusing  sdNLS equation.
(a) one-soliton solution for $a_0=0.4, k_1=0.2i, c_1=1, d_1=-i$.
(b) one-soliton solution for $a_0=0.4, k_1=0.2i, c_1=1, d_1=i$.
(c) one-soliton solution for $a_0=0.4, k_1=0.2i, c_1=1, d_1=1-i$.
(d) one-soliton solution for $a_0=0.4, k_1=0.2i, c_1=-d_1=1$.
(e) one-soliton solution for $a_0=0.4, k_1=0.2i, c_1=d_1=1$.
(f) doubly periodic solution for $a_0=0.1, k_1=0.2i, c_1=d_1=1$.
}
\end{figure}

We are interested in two special cases of the above one soliton solution.
The first  case is for
\begin{eqnarray}
k_1\in i\mathbb R,~~~ 4a_0^2>-(e^{k_1}-e^{-k_1})^2,~~~a_0^2<1,
\end{eqnarray}
which yields real $\lambda$ and $\eta$, i.e. $b_1=b_2=0$.
It follows that
\begin{eqnarray}\label{Qn-1}
    Q_n=\frac{D_1e^{2a_2t}+D_2e^{-2a_2t}+D_3e^{2a_1n}+D_4e^{-2a_1n}}
    {C_1e^{2a_2t}+C_2e^{-2a_2t}+C_3e^{2a_1n}+C_4e^{-2a_1n}}e^{2i a_0^2 t}.
\end{eqnarray}
Note that in this case we have $\xi(-k_1)=(\xi(k_1))^*$, $|\xi(k_1)|=1$ and
\begin{equation*}
    \frac{D_1}{C_1}=a_0(\xi^2(k_1))^*, ~~~ \frac{D_2}{C_2}=a_0 \xi^2(k_1), ~~~ \frac{D_3}{C_3}=\frac{D_4}{C_4}=-a_0.
\end{equation*}
Then, it is easy to see that
for any fixed $n$ we have
\[
    \lim_{t\rightarrow \pm\infty}|Q_n|^2 =a_0^2,
\]
and for any fixed $t$  we have
\[
    \lim_{n\rightarrow \pm\infty}|Q_n|^2 =a_0^2.
\]
This indicates $|Q_n|^2$   asymptotically lives on the plane $|Q_n|^2=a_0^2$.
The illustration in Fig.\ref{Fig-5a} shows that the squared envelop $|Q_n|^2$ of the one-soliton solution
behaves like interaction of two solitons.

To see more asymptotic property of $Q_n$,
we assume  $a_1>0,~a_2>0$ (other situations can be analyzed similarly)
introduce
\begin{equation}\label{X1X2}
X_1=a_1 n+a_2 t,~~~ X_2=a_1n-a_2 t,
\end{equation}
and rewrite \eqref{Qn-1} as
\begin{equation}\label{Qn-2}
    Q_n=\frac{D_1e^{4a_2t}+D_2 +D_3e^{2X_1}+D_4e^{-2X_2}}
    {C_1e^{4a_2t}+C_2 +C_3e^{2X_1}+C_4e^{-2X_2}}e^{2i a_0^2 t}.
\end{equation}
Considering the above $Q_n$ in the coordinate frame $(X_1,t)$, i.e.
\begin{equation}\label{Qn-3}
    Q_n=\frac{D_1e^{4a_2t}+D_2 +D_3e^{2X_1}+D_4e^{-2X_1+4a_2 t}}
    {C_1e^{4a_2t}+C_2 +C_3e^{2X_1}+C_4e^{-2X_1+4a_2 t}}e^{2i a_0^2 t},
\end{equation}
we find that
\begin{equation}\label{Qn-4}
    Q_n\sim \left\{
    \begin{array}{ll}
    q_{1,n}^{+}\doteq \frac{D_4 +D_1e^{2X_1 }}
    {C_4 +C_1e^{2X_1 }}e^{2i a_0^2 t},&  t\to +\infty,\\
    q_{1,n}^{-}\doteq \frac{ D_2 +D_3e^{2X_1} }
    { C_2 +C_3e^{2X_1} }e^{2i a_0^2 t},&  t\to -\infty.
    \end{array}\right.
\end{equation}
For convenience, we introduce notations
\begin{equation*}
\xi(k_1)=e^{ i\theta_1},~~
\mu=\left|\frac{c^*_1e^{-\lambda}\xi(k_1)}{d^*_1}\right|,~~
\nu=\left|\frac{e^\lambda+e^{-\lambda}}{(\xi(k_1))^*-\xi(k)}\right|, ~~
\frac{1}{\mu\nu}\frac{c^*_1e^{-\lambda}\xi(k_1)}{d^*_1}
\frac{e^\lambda+e^{-\lambda}}{(\xi(k))^*-\xi(k)} = e^{i\theta_2},
\end{equation*}
where $\theta_1, \theta_2, \mu, \nu\in \mathbb{R}$.
Then we have
\[|q_{1,n}^{\pm}|^2=a_0^2\frac{1+ \mu^2\nu^{\mp 2} y_1^2 -\mu\nu^{\mp 1} y_1(e^{i(2\theta_1+\theta_2)}
+e^{-i(2\theta_1+\theta_2)})}
{1+\mu^2\nu^{\mp 2} y_1^2+\mu\nu^{\mp 1} y_1(e^{i\theta_2}+e^{-i\theta_2})},~~ ~ (y_1=e^{2X_1}).\]
$|q_{1,n}^{+}|^2$ and $|q_{1,n}^{-}|^2$  describe a same soliton
(we call it $X_1$-soliton for convenience)
living on the background $|Q_n|^2=a_0^2$
and characterized by the following features:
\begin{align*}
&{\rm trajectory}: ~X_1=-\frac{1}{2}\ln(\mu \nu^{\mp 1}),\\
&{\rm velocity:} ~n^\prime(t)=-\frac{a_2}{a_1},\\
&{\rm amplitude:} ~ A_1=a_0^2\frac{1-\Re(e^{i(2\theta_1+\theta_2)})}{1+\Re(e^{i\theta_2})}.
\end{align*}
This indicates that the $X_1$-soliton obtains a phase shift $\ln(\nu)$ after interaction
with another soliton (which we call  $X_2$-soliton for convenience, see the following).

Considering   $Q_n$ \eqref{Qn-1} in the coordinate frame $(X_2,t)$, i.e.
\begin{equation}\label{Qn-5}
    Q_n=\frac{D_1+D_2 e^{-4a_2t}+D_3e^{2X_2}+D_4e^{-2X_2-4a_2t}}
    {C_1+C_2e^{-4a_2t} +C_3e^{2X_2}+C_4e^{-2X_2-4a_2t}}e^{2i a_0^2 t},
\end{equation}
which yields
\begin{equation}\label{Qn-6}
    Q_n\sim \left\{
    \begin{array}{ll}
    q_{2,n}^{+}\doteq \frac{D_1 +D_3 e^{2X_2 }}
    {C_1 +C_3 e^{2X_2 }}e^{2i a_0^2 t}, &  t\to +\infty,\\
    q_{2,n}^{-}\doteq \frac{ D_4 +D_2e^{2X_2} }
    { C_4 +C_2e^{2X_1} }e^{2i a_0^2 t},&  t\to -\infty,
    \end{array}\right.
\end{equation}
and
\[|q_{2,n}^{\pm}|^2=a_0^2\frac{1+ \mu^{-2}\nu^{\pm 2} y_2^2 -\mu^{-1}\nu^{\pm 1}
y_2(e^{i(2\theta_1-\theta_2)}+e^{-i(2\theta_1-\theta_2)})}
{1+\mu^{-2}\nu^{\pm 2} y_2^2+\mu^{-1}\nu^{\pm 1} y_2(e^{i\theta_2}+e^{-i\theta_2})},~~ ~ (y_2=e^{2X_2}).\]
Both $|q_{2,n}^{+}|^2$ and $|q_{2,n}^{-}|^2$  describe the $X_2$-soliton
living on the background $|Q_n|^2=a_0^2$
and characterized by:
\begin{align*}
&{\rm trajectory}: ~X_2=-\frac{1}{2}\ln(\mu^{-1} \nu^{\pm 1}),\\
&{\rm velocity:} ~n^\prime(t)= \frac{a_2}{a_1},\\
&{\rm amplitude:} ~ A_2=a_0^2\frac{1-\Re(e^{i(2\theta_1-\theta_2)})}{1+\Re(e^{i\theta_2})}.
\end{align*}
The phase shift of the $X_2$-soliton due to interaction with the $X_1$-soliton is  $\ln(\nu)$ as well.

Now, the interaction depicted in Fig.\ref{Fig-5a} can be well understood.
In addition, the asymptotic amplitudes of the  $X_1$-soliton and the  $X_2$-soliton
show that the amplitudes can be either larger or less than $a_0^2$ or equal to $a_0^2$,
depending on the values of $\Re(e^{i(2\theta_1\pm \theta_2)})$ and $\Re(e^{i\theta_2})$.
This allows us to have various types of interactions, such as bright-bright, dark-dark, bright-dark
and just a single bright soliton and a single dark soliton. Their illustration are given in
Fig.\ref{Fig-5a}-\ref{Fig-5e}.
It is also predictable the various interactions of two-soliton solutions,
while in this paper  we skip presenting their formulae and illustrations.

In the second special case of our interest, we consider
\begin{eqnarray}
    k_1\in i\mathbb R,~~~ 4a_0^2<-(e^{k_1}-e^{-k_1})^2,~~~a_0^2<1,
\end{eqnarray}
which  yields purely imaginary $\lambda$ and $\eta$, i.e. $a_1=a_2=0$.
The solution of this case reads
\begin{eqnarray}
    Q_n=\frac{D_1e^{2ib_1n}+D_2e^{-2ib_1n}+D_3e^{2ib_2t}+D_4e^{-2ib_2t}}
    {C_1e^{2ib_1n}+C_2e^{-2ib_1n}+C_3e^{2ib_2t}+C_4e^{-2ib_2t}}e^{2ia_0^2t}.
\end{eqnarray}
This demonstrates that $|Q_n|^2$ is a doubly period function
with period $T_1=\pi/b_1$ in $n$-direction and  period $T_2=\pi/b_2$ in $m$-direction,
see Fig.\ref{Fig-5f} as an illustration.

\captionsetup[figure]{labelfont={bf},name={Fig.},labelsep=period}

\subsubsection{Rational solutions}\label{sec-5-2-2}

Similar to the classical case,  we should assume $0<a_0<1$, introduce a parameter $\kappa$ by
\begin{equation}\label{nonlocal-kappa}
\kappa={\sqrt{(e^k-e^{-k})^2+4a_0^2}},
\end{equation}
and express the involved elements in  $\phi_n$ and $\psi_n$ defined in \eqref{phipsi} as:
\begin{subequations}
\begin{align}
& e^k=\frac{1}{2}\sqrt{\kappa^2+4-4a_0^2}+\frac12\sqrt{\kappa^2-4a^2_0},~~~
e^{-k}= \frac{1}{2}\sqrt{\kappa^2+4-4a_0^2}-\frac12\sqrt{\kappa^2-4a^2_0},\\
&  e^{\lambda}=\frac{\kappa+\sqrt{\kappa^2+4-4a_0^2}}{\sqrt{4-4a_0^2}}, ~~~ e^{-\lambda}=\frac{-\kappa+\sqrt{\kappa^2+4-4a_0^2}}{\sqrt{4-4a_0^2}},\\
&    \eta=-\frac i2\kappa\sqrt{\kappa-4a_0^2},\\
&    \xi(k)=\tilde{\xi}(\kappa)=\frac{1}{2a_0}\Bigl(\sqrt{\kappa^2-4a_0^2}-\kappa\Bigr),~~~
\xi(-k)=-\tilde{\xi}(-\kappa).
\end{align}
\end{subequations}
We also assume $c$ and $d$ are functions of $\kappa$ and satisfy
\begin{equation}\label{nonlocal-cd-kappa}
c(\kappa)=-d(-\kappa).
\end{equation}
Thus,  $\phi_n$ and $\psi_n$  in \eqref{phipsi} can be written in terms of $\kappa$ as
\begin{subequations}\label{nonlocal-phipsi-kappa}
\begin{align}
&\phi_n =\Biggl[ c(\kappa)
\Biggl(\frac{\kappa+\sqrt{\kappa^2+4-4a_0^2}}{\sqrt{4-4a_0^2}}\Biggr)^n e^{\eta t}
-c(-\kappa) \Biggl(\frac{\kappa+\sqrt{\kappa^2+4-4a_0^2}}{\sqrt{4-4a_0^2}}\Biggr)^{-n} e^{-\eta t}
\Biggr] e^{- i a^2_0  t},\\
&\psi_n =\Biggl[-c(\kappa) \tilde{\xi}(\kappa) \Biggl(\frac{\kappa+\sqrt{\kappa^2+4-4a_0^2}}{\sqrt{4-4a_0^2}}\Biggr)^n e^{\eta t}
+c(-\kappa)\tilde{\xi}(-\kappa)
\Biggl(\frac{\kappa+\sqrt{\kappa^2+4-4a_0^2}}{\sqrt{4-4a_0^2}}\Biggr)^{-n} e^{-\eta t}
\Biggr] e^{ i a^2_0 t}.
\end{align}
\end{subequations}
In addition, we assume   $c(\kappa)$ and $d(\kappa)$
to be the following series (which agree with \eqref{nonlocal-cd-kappa}):
\begin{eqnarray}
    c(\kappa)=\sum_{j=0}^\infty s_j\kappa^j,~~ d(\kappa)=\sum_{j=0}^\infty (-1)^{j+1}s_j\kappa^j,
\end{eqnarray}
where $\{s_j\}$ are arbitrary complex numbers.

Then, we can have expansions for $\phi_n$ and $\psi_n$:
\begin{eqnarray}
 \phi_n=\sum_{j=0}^\infty R_{2j+1,n}\kappa^{2j+1}, && \psi_n=\sum_{j=0}^\infty S_{2j+1,n}\kappa^{2j+1},
\end{eqnarray}
where
\begin{equation}
R_{2j+1,n}=\frac{\partial_{\kappa}^{2j+1}}{(2j+1)!}\phi_n|_{\kappa=0}, ~~
S_{2j+1,n}=\frac{\partial_{\kappa}^{2j+1}}{(2j+1)!}\psi_n|_{\kappa=0}.
\end{equation}
Introducing
\begin{eqnarray}
    \Phi_n=(R_{1,n},R_{3,n},\cdots, R_{2m+1,n}, S_{1,1-n}^*,S_{3,1-n}^*,\cdots, S_{2m+1,1-n}^*)^T
\end{eqnarray}
and denoting $e^{\mathbf{K}_{m+1}}$ as in \eqref{5.21} and \eqref{5.22},
one can prove that such a $\Phi_n$ satisfies the equation set \eqref{sec3-phipsi-reduction2-case1}
where
\begin{eqnarray*}
 \delta=1,& A=\mathrm{Diag}(e^{\mathbf{K}_{m+1}},e^{\mathbf{K}_{m+1}^*}), &
 T=\left(\begin{matrix}
    \mathbf{0}_{m+1} & I_{m+1}\\
    -I_{m+1} & \mathbf{0}_{m+1}
\end{matrix}\right).
\end{eqnarray*} 
Thus, explicit rational solution for the reverse-space defocusing sdNLS equation is expressed as
\begin{subequations}\label{nonlocal-rw-solu}
\begin{equation}
Q_n=\frac{G_n}{F_n},
\end{equation}
where $F_n$ and $G_n$ are the quasi double Casoratians composed by the above
$\Phi_n, A$ and $T$:
\begin{align}
    F_n=& |A\Phi_n,A^3\Phi_n,\cdots,A^{2m+1}\Phi_n;
    T\Phi_{1-n}^*,A^{2}T\Phi_{1-n}^*,\cdots,A^{2m}T\Phi_{1-n}^*|, \label{nonlocal-Fn-RW}\\
    G_n=& |\Phi_n, A^2 \Phi_{n}, \cdots,A^{2m+2}\Phi_{n};
     AT\Phi_{1-n}^*, A^{3}T\Phi_{1-n}^*,\cdots, A^{2m-1}T \Phi_{1-n}^*| \nonumber\\
    &+ (-1)^m q_n  |\Phi_n, A^2 \Phi_{n}, \cdots,A^{2m}\Phi_{n};
    AT\Phi_{1-n}^*, A^{3}T\Phi_{1-n}^*,\cdots, A^{2m+1}T \Phi_{1-n}^*|.
\end{align}
\end{subequations}

In the following we just show the simplest rational solution which is resulted from $m=0$ and $0<a_0<1$.
In this case we get the first order rational solution
\begin{subequations}
\begin{equation}\label{Qn-7}
Q_n=-a_0\Bigr(1-\frac{W_n}{U_n}\Bigr)e^{2ia^2_0t}
\end{equation}
where
\begin{align}
W_n=& i\frac{|s_0|^2}{a_0}\left[8a_0t+8\Re\Bigl(\frac{s_1}{s_0}\Bigr)
+\frac{1}{\sqrt{1-a_0^2}}\biggl(1-i\frac{\sqrt{1-a_0^2}}{a_0}\biggr)^2\right]\\
U_n=&|s_0|^2\Biggr[\biggl(\frac{1}{\sqrt{1-a_0^2}}+2a_0t+2\Re\Bigl(\frac{s_1}{s_0}\Bigr)\biggr)^2
+\Bigl(2a_0t+2\Re\Bigl(\frac{s_1}{s_0}\Bigr)\Bigr)^2- \frac{2 n^2}{ 1-a_0^2 } \nonumber\\
&+\Bigl(2\Im\Bigl(\frac{s_1}{s_0}\Bigr)+\frac{1}{a_0}\Bigr)^2
+4\Im\Bigl(\frac{s_1}{s_0}\Bigr)^2-2i\Bigl(4\Im\Bigl(\frac{s_1}{s_0}\Bigr)+\frac{1}{a_0}\Bigr)
\frac{n}{\sqrt{1-a_0^2}}\Biggr].
\end{align}
\end{subequations}

To see more insights about the waves described by such a rational solution,
we introduce
\[X_1=2a_0t+\frac{n}{\sqrt{1-a_0^2}},~~ X_2=2a_0t-\frac{n}{\sqrt{1-a_0^2}}.\]
Then, we write $U_n$ and $W_n$ in terms of $X_1$ and $X_2$, i.e.
\begin{align*}
U_n=&|s_0|^2\Biggl[\Bigl(X_1+\frac{1}{\sqrt{1-a_0^2}}+2\Re\Bigl(\frac{s_1}{s_0}\Bigr)\Bigr)
\Bigl(X_2+\frac{1}{\sqrt{1-a_0^2}}+2\Re\Bigl(\frac{s_1}{s_0}\Bigr)\Bigr)\\
&+\Bigl(X_1+2\Re\Bigl(\frac{s_1}{s_0}\Bigr)\Bigr)\Bigl(X_2+2\Re\Bigl(\frac{s_1}{s_0}\Bigr)\Bigr)\\
&+\Bigl(2\Im\Bigl(\frac{s_1}{s_0}\Bigr)+\frac{1}{a_0}\Bigr)^2
+4\Im\Bigl(\frac{s_1}{s_0}\Bigr)^2-i\Bigl(4\Im\Bigl(\frac{s_1}{s_0}\Bigr)+\frac{1}{a_0}\Bigr)(X_2-X_1)
\Biggr],\\
W_n=& i\frac{|s_0|^2}{a_0}\Biggl[ 2X_1+2X_2+8\Re\Bigl(\frac{s_1}{s_0}\Bigr)
     +\frac{1}{\sqrt{1-a_0^2}}\biggl(1-i\frac{\sqrt{1-a_0^2}}{a_0}\biggr)^2\Biggr],
\end{align*}
and consider $Q_n$ in the coordinate systems $(X_1,t)$ and $(X_2,t)$ respectively.
By taking $t\to \pm\infty$, we can obtain asymptotic feathers for  $Q_n$.
It turns out that, in the coordinate frame $(X_1,t)$, we have
\begin{equation*}
 |Q_n|^2\sim a_0^2\frac{\Bigl(2a_0X_1+\frac{a_0}{\sqrt{1-a_0^2}}+4a_0\Re(\frac{s_1}{s_0})\Bigr)^2
 +(4a_0\Im(\frac{s_1}{s_0})+3)^2}
 {\Bigl(2a_0X_1+\frac{a_0}{\sqrt{1-a_0^2}}+4a_0\Re(\frac{s_1}{s_0})\Bigr)^2
 +(4a_0\Im(\frac{s_1}{s_0})+1)^2},~~~ (t\to \pm\infty),
\end{equation*}
which is an algebraic soliton (we call it $X_1$-soliton) traveling with:
\begin{align*}
&{\rm trajectory}: ~X_1=-\frac{1}{\sqrt{1-a_0^2}}-4\Re\Bigl(\frac{s_1}{s_0}\Bigr),\\
&{\rm velocity:} ~n^\prime(t)=\frac{-2a_0}{\sqrt{1-a_0^2}},\\
&{\rm amplitude:} ~ A_1=a_0^2\frac{(4a_0\Im(\frac{s_1}{s_0})+3))^2}{(4a_0\Im(\frac{s_1}{s_0})+1))^2}.
\end{align*}
In a similar way, in the coordinate frame $(X_2,t)$, we get
\begin{equation*}
 |Q_n|^2\sim a_0^2\frac{\Bigl( 2a_0X_2+\frac{a_0}{\sqrt{1-a_0^2}}+4a_0\Re(\frac{s_1}{s_0})\Bigr)^2
 +(4a_0\Im(\frac{s_1}{s_0})-1)^2}
 {\Bigl(2a_0X_2+\frac{a_0}{\sqrt{1-a_0^2}} +4a_0\Re(\frac{s_1}{s_0})\Bigr)^2
 +(4a_0\Im(\frac{s_1}{s_0})+1)^2},~~~ (t\to \pm\infty),
\end{equation*}
which described the  $X_2$-algebraic soliton traveling with:
\begin{align*}
&{\rm trajectory}: ~X_2=-\frac{1}{\sqrt{1-a_0^2}}-4\Re\Bigl(\frac{s_1}{s_0}\Bigr),\\
&{\rm velocity:} ~n^\prime(t)=\frac{2a_0}{\sqrt{1-a_0^2}},\\
&{\rm amplitude:} ~ A_2=a_0^2\frac{(4a_0\Im(\frac{s_1}{s_0})-1))^2}{(4a_0\Im(\frac{s_1}{s_0})+1))^2}.
\end{align*}

\begin{figure}[ht]
\centering
\subfigure[ ]{
\begin{minipage}[t]{0.32\linewidth}
\centering
\includegraphics[width=2.0in]{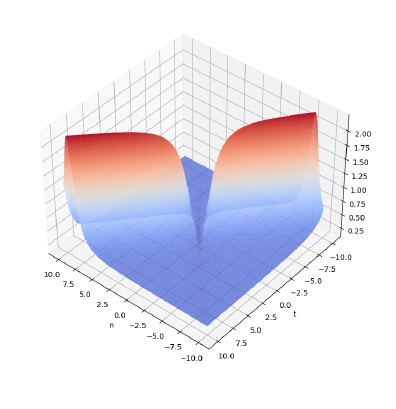}
\end{minipage}\label{nonlocal-rational-1}
}%
\subfigure[ ]{
\begin{minipage}[t]{0.32\linewidth}
\centering
\includegraphics[width=2.0in]{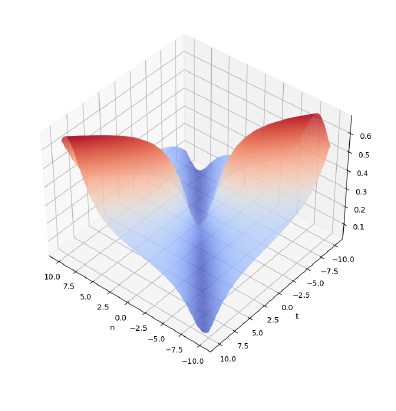}
\end{minipage}\label{nonlocal-rational-2}
}%
\subfigure[ ]{
\begin{minipage}[t]{0.32\linewidth}
\centering
\includegraphics[width=2.0in]{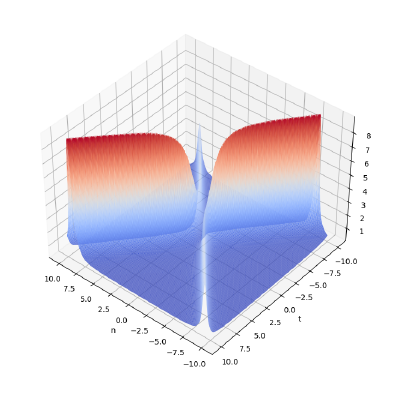}
\end{minipage}\label{nonlocal-rational-3}
}
\caption{\label{Fig-6} Shape and motion of the squared envelop
 of the rational solutions of the reverse-space defocusing sdNLS equation.
(a) algebraic soliton solution for $a_0=0.5, c(\kappa)=-d(-\kappa)=1$.
(b) algebraic soliton solution for $a_0=0.5, c(\kappa)=-d(-\kappa)=1+i\kappa$.
(c) algebraic soliton solution for $a_0=0.5, c(\kappa)=-d(-\kappa)=1-0.3i\kappa$.
}
\end{figure}

These analyses are helpful in understanding dynamics of the algebraic solitons, for example,
both $X_1$-soliton and $X_2$-soliton gain zero shift after interaction. In addition, it is possible to have
a single bright soliton (e.g. $A_1>a_0^2, A_2=a_0^2$).
The squared envelop are depicted in Fig.\ref{Fig-6}.
One can also analyze second-order algebraic solitons with illustrations
 but we skip them in this paper.

\captionsetup[figure]{labelfont={bf},name={Fig.},labelsep=period}

\subsection{The reverse-space  focusing sdNLS with a hyperbolic wave background}\label{sec-5-3}

In this subsection we consider the reverse-space focusing sdNLS equation \eqref{sec2-dNLS-equ3} ($\delta=-1$)
with the following background (cf.\cite{XALX-SAPM-2024})
\begin{eqnarray}\label{qn}
    q_n=a_0\tanh(\mu n+i\mu\omega  )e^{2ia_0^2t}, ~~ r_n=-q^*_{-n},
\end{eqnarray}
where  $\mu,\omega\in\mathbb{R}$ and $a_0=\tanh(\mu)$.
Before we proceed, one should recall the notations and formulae given in
Sec.\ref{sec-4-3-2}.

When $m=0$ and $\mathbf K_1=k_1$, we have
\begin{equation}
F_n=\alpha_n^{1/2}\begin{vmatrix}
        \Phi^+_{1+n} &\Psi^+_n\\
        (\Psi^+_{-n})^* &(\Phi_{1-n}^+)^*
    \end{vmatrix},~~~
G_n=\alpha_n^{1/2}\begin{vmatrix}
    \Phi^+_{n} &e^{k_1}\Phi^+_{1+n}\\
    (\Psi^+_{1-n})^* &(-e^{k_1^*}\Psi^+_{-n})^*
\end{vmatrix} ~~,
\end{equation}
where
\[ \Phi^+_{n}=\hat\gamma_{n}\hat\phi_n(k_1,c_1,d_1), ~~
\Psi^+_{n}=\hat\gamma_{n}\hat\psi_n(k_1,c_1,d_1), \]
$\hat\phi_n(k_1,c_1,d_1)$ and $\hat\psi_n(k_1,c_1,d_1)$ are given in \eqref{4.38c} and \eqref{4.38d},
$\alpha_n$ and $\hat\gamma_{n}$ are defined in \eqref{4.38b}, while here they are
\begin{eqnarray}
 \alpha_n=1-a_0^2\tanh^2(\mu n+i\mu\omega ),~~ ~
 \hat\gamma_{n}=\prod^{n-1}_{s=-\infty}\sqrt{\frac{1-a_0^2}{\alpha_s}}.
\end{eqnarray}
In practice, one can write $F_n$ and $G_n$ into
\begin{equation}\label{5.63}
F_n=\alpha_n^{1/2}\hat\gamma_n\hat\gamma^*_{-n} \hat F_n,~~
G_n=\alpha_n^{1/2}\hat\gamma_n\hat\gamma^*_{-n} \hat G_n,
\end{equation}
where
\begin{equation}
\hat F_n=\begin{vmatrix}
        \gamma_n\hat\Phi^+_{1+n} &\hat\Psi^+_n\\
        (\hat\Psi^+_{-n})^* &\gamma_n(\hat\Psi_{1-n}^+)^*
    \end{vmatrix},~~~
\hat G_n=\begin{vmatrix}
        \hat\Phi^+_n & \gamma_ne^{k_1}\hat\Phi^+_{1+n} \\
         \gamma_n(\hat\Psi^+_{1-n})^* &(e^{k_1}\hat\Psi^+_{-n})^*
    \end{vmatrix},
\end{equation}
and
\[\gamma_n=\frac{ \hat\gamma_{n+1}}{ \hat\gamma_{n}}
=\sqrt\frac{1-a_0^2}{\alpha_n}
=\sqrt\frac{1-a_0^2}{1-a_0^2\tanh^2(\mu n+i\mu\omega )}.
\]

Apparently,
\begin{eqnarray}
    Q_n=\frac{G_n}{F_n}=\frac{\hat G_n}{\hat F_n}.
\end{eqnarray}
In practice, we calculate $Q_n$ using $\hat F_n$ and $\hat G_n$,  of which the explicit forms are
%\begin{subequations}
\begin{align*}
& \hat F_n  =C_{1,n}e^{2(a_2t + i b_1n)}+C_{2,n}e^{-2(a_2t + i b_1n)}
    +C_{3,n}e^{2(a_1n+b_2it)}+C_{4,n}e^{-2(a_1n+ib_2t)},\\
& \hat G_n=[D_{1,n}e^{2(a_2t + i b_1n)}+D_{2,n}e^{-2(a_2t + i b_1n)}
+D_{3,n}e^{2(a_1n+ib_2t)}+D_{4,n} e^{-2(a_1n+ib_2t)}]e^{2ia_0^2t},
\end{align*}
%\end{subequations}
where
\begin{align*}
C_{1,n}=&|c_1|^2\Bigl[\gamma_n^2e^{\lambda+\lambda^*}
           \Big(\xi(-k_1)e^{k_1}+\tanh(\mu n+i\mu \omega)\Big)
           \Big((\xi(-{k_1}))^*e^{{k_1^*}}-\tanh(\mu n+i\mu \omega)\Big)\\
   & -\Big(\xi({k_1})+e^{k_1}\tanh(\mu n-\mu+i \mu \omega )\Big)
  \Bigl ((\xi({k_1}))^*-e^{{k_1^*}}\tanh(\mu n+\mu +i\mu \omega )\Bigr)\Bigr],\\
C_{2,n}=&|d_1|^2\Bigr[\gamma_n^2e^{-(\lambda+\lambda^*)}
     \Big (\xi({k_1})e^{k_1}-\tanh(\mu n+i \mu \omega )\Big)
     \Big((\xi({k_1}))^* e^{{k_1^*}}+\tanh(\mu n+i\mu \omega )\Big)\\
    & -\Big(\xi(-{k_1})-e^{k_1}\tanh(\mu n-\mu+ i\mu \omega)\Big)
    \Big((\xi(-{k_1}))^* +e^{{k_1^*}}\tanh(\mu n+\mu+ i\mu \omega )\Big)\Bigr],\\
C_{3,n}=& c_1d_1^*\Bigr[\gamma_n^2e^{\lambda-\lambda^*}
              \Big(\xi(-{k_1})e^{k_1}+\tanh(\mu n+i\mu \omega )\Big)
              \Big(-(\xi({k_1}))^*e^{{k_1^*}}-\tanh(\mu n+i\mu \omega )\Bigr)\\
    &+\Big(\xi({k_1})+e^{k_1}\tanh(\mu n-\mu+i\mu \omega )\Big)
    \Big((\xi(-{k_1}))^*+ e^{{k_1^*}}\tanh(\mu n+\mu+i\mu \omega )\Big) \Bigr],\\
C_{4,n}=&c_1^*d_1\Bigr[\gamma_n^2e^{-(\lambda-\lambda^*)}
            \Big (\xi^*(-{k_1})e^{{k_1^*}}-\tanh(\mu n+i\mu \omega )\Big)
            \Big(-\xi({k_1})e^{k_1}+\tanh(\mu n+i \mu \omega )\Big)\\
    &+\Big((\xi({k_1}))^*-e^{{k_1^*}}\tanh(\mu n+\mu+i\mu \omega )\Big)
    \Big(\xi(-{k_1})-e^{k_1}\tanh(\mu n-\mu+i\mu \omega )\Big) \Bigr],\\
D_{1,n}=& |c_1|^2\Bigr[\gamma_n^2e^{k_1}e^{\lambda+\lambda^*}
             \Big(\xi(-{k_1})e^{k_1}+\tanh(\mu n+i\mu \omega )\Big)
             \Big((\xi({k_1}))^*-e^{{k_1^*}}\tanh(\mu n+i\mu \omega )\Big)\\
    &-e^{{k_1^*}}\Big(\xi(-{k_1})e^{k_1}+\tanh(\mu n-\mu+i \mu \omega)\Big)
       \Big((\xi({k_1}))^*-e^{{k_1^*}}\tanh(\mu n+\mu+i\mu \omega)\Big)\Bigr],\\
D_{2,n}=& |d_1|^2\Bigr[\gamma_n^2e^{k_1}e^{-(\lambda+\lambda^*)}
             \Big(\xi({k_1})e^{k_1}-\tanh(\mu n+i\mu \omega)\Big)
             \Big((\xi(-{k_1}))^*+e^{{k_1^*}}\tanh(\mu n+i\mu \omega)\Big)\\
    & -e^{{k_1^*}}\Big(\xi({k_1})e^{k_1}-\tanh(\mu n-\mu+i\mu \omega )\Big)
       \Big ((\xi(-{k_1}))^* +e^{{k_1^*}}\tanh(\mu n+\mu+i\mu \omega )\Big)\Bigr],\\
D_{3,n}=& c_1d_1^*\Bigr[-\gamma_n^2e^{k_1}e^{\lambda-\lambda^*}
              \Big (\xi(-{k_1})e^{k_1}+\tanh(\mu n+i\mu \omega )\Big)
              \Big((\xi(-{k_1}))^* +e^{{k_1^*}}\tanh(\mu n+i\mu \omega)\Big)\\
    & +e^{{k_1^*}}\Big(\xi(-{k_1})e^{k_1}+\tanh(\mu n-\mu+i\mu \omega)\Big)
       \Big ((\xi(-{k_1}))^* +e^{{k_1^*}}\tanh(\mu n+\mu+i\mu \omega )\Big)\Bigr],\\
D_{4,n}=& c_1^*d_1\Bigr[-\gamma_n^2e^{k_1}e^{-(\lambda-\lambda^*)}
          \Big(\xi({k_1})e^{k_1}-\tanh(\mu n+i \mu \omega )\Big)
          \Big((\xi({k_1}))^*-e^{{k_1^*}}\tanh(\mu n+i\mu \omega )\Big)\\
    & +e^{{k_1^*}}\Big(\xi({k_1})e^{k_1}-\tanh(\mu n-\mu+i\mu \omega )\Big)
       \Big ((\xi({k_1}))^*-e^{{k_1^*}}\tanh(\mu n+\mu+i\mu \omega )\Big)\Bigr],
\end{align*}
and we have taken $\lambda=a_1+ib_1,~\eta=a_2+ib_2$.

Note that, although parameters $C_{j,n},~ D_{j,n},~ j=1,2,3,4$ are functions of $n$,
they tend to finite-valued constants when $n\rightarrow \pm \infty$.
It is then possible to analyze the asymptotic behavior of the squared envelope $|Q_n|^2$.
However, in the following we only present two interesting cases of one-soliton solutions and their illustrations.

\begin{figure}[ht]
\centering
\subfigure[ ]{\label{Fig-7a}
\begin{minipage}[t]{0.32\linewidth}
\centering
\includegraphics[width=2.0in]{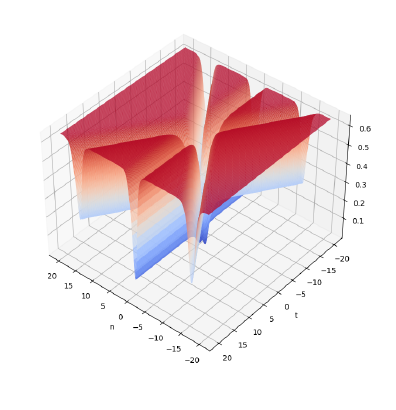}
\end{minipage}%
}%
\subfigure[ ]{\label{Fig-7b}
\begin{minipage}[t]{0.32\linewidth}
\centering
\includegraphics[width=2.0in]{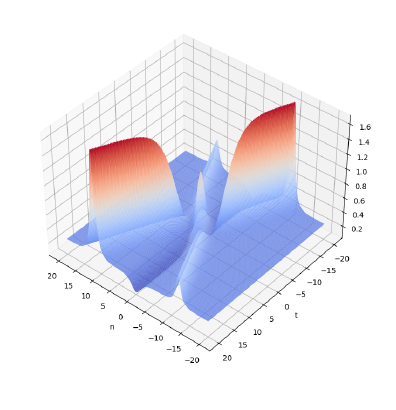}
\end{minipage}%
}
\subfigure[ ]{\label{Fig-7c}
\begin{minipage}[t]{0.32\linewidth}
\centering
\includegraphics[width=2.0in]{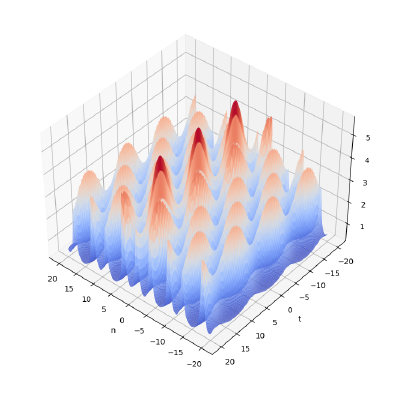}
\end{minipage}%
}%

\caption{ Shape and motion of the squared envelop of the solutions of the reverse-space focusing sdNLS equation.
(a) one-soliton solution for $a_0=0.8, \omega=4, k_1=0.5i, c_1=d_1=1$.
(b) one-soliton solution for $a_0=0.5, \omega=2, k_1=0.25i, c_1=1, d=-1+i$.
(c) periodic solution for $a_0=0.2, \omega=0, k_1=0.4i, c_1=1, d=1+i$.
}\label{Fig-7}
\end{figure}

The first case is that
\begin{eqnarray}
    k_1\in i\mathbb R, ~~ (e^{k_1}-e^{-k_1})^2+a_0^2>0.
\end{eqnarray}
It then follows that $ b_1=b_2=0$ and
\begin{eqnarray}
    Q_n=\frac{D_{1,n}e^{2a_2t}+D_{2,n}e^{-2a_2t}+D_{3,n}e^{2a_1n}+D_{4,n}e^{-2a_1n}}
    {C_{1,n}e^{2a_2t}+C_{2,n}e^{-2a_2t}+C_{3,n}e^{2a_1n}+C_{4,n}e^{-2a_1n}}e^{2ia_0^2t}.
\end{eqnarray}
The illustrations of the resulted $|Q_n|^2$ are given in Fig.\ref{Fig-7a} and \ref{Fig-7b}.
Note that the wave at $n=0$ is due to the background $|q_n|^2$
with $q_n$ given in \eqref{qn}.

The second case is that
\begin{eqnarray}
    k_1\in i\mathbb R, ~~ (e^{k_1}-e^{-k_1})^2+a_0^2<0,
\end{eqnarray}
which yields $a_1=a_2=0$ and
\begin{eqnarray}
    Q_n=\frac{D_{1,n}e^{2ib_1n}+D_{2,n}e^{-2ib_1n}+D_{3,n}e^{2ib_2t}+D_{4,n}e^{-2ib_2t}}
    {C_{1,n}e^{2ib_1n}+C_{2,n}e^{-2ib_1n}+C_{3,n}e^{2ib_2t}+C_{4,n}e^{-2ib_2t}}e^{2ia_0^2t}.
\end{eqnarray}
Apparently, this generates a periodic $|Q_n|^2$ with a period $T_2=\pi/b_2$ in $t$-direction.
Note that in $n$-direction $|Q_n|^2$ is not periodic (but is quasi-periodic) due to
the non-periodic background  $|q_n|^2$.
Such a wave is depicted in Fig.\ref{Fig-7c}.

\captionsetup[figure]{labelfont={bf},name={Fig.},labelsep=period}

Finally, we note that the expressions in \eqref{5.63} hold for a general $m$ and $\mathbf{K}_{m+1}$, where
{\small
\begin{align*}
&\hat F_n=\begin{vmatrix}
        \gamma_n\hat\Phi^+_{1+n}& \cdots &\gamma_ne^{2m\mathbf{K}_{m+1}}\hat\Phi^+_{1+n}
        &\hat\Psi^+_{n}
        & \cdots & e^{2m\mathbf{K}_{m+1}}\hat\Psi^+_{n}\\
        (\hat\Psi^+_{-n})^*& \cdots &(e^{2m\mathbf{K}_{m+1}}\hat\Psi^+_{-n})^*
        &\gamma_n(\hat\Phi^+_{1-n})^*
        & \cdots &\gamma_n(e^{2m\mathbf{K}_{m+1}}\hat\Phi^+_{1-n})^*
    \end{vmatrix},\\
&\hat G_n= \begin{vmatrix}
        \hat\Phi^+_n &\gamma_ne^{\mathbf{K}_{m+1}}\hat\Phi^+_{1+n}& \cdots &\gamma_ne^{(2m+1)\mathbf{K}_{m+1}}\hat\Phi^+_{1+n}
        &e^{\mathbf{K}_{m+1}}\hat\Psi^+_{n}& \cdots &e^{(2m-1)\mathbf{K}_{m+1}}\hat\Psi^+_{n}\\
        \gamma_n(\hat\Psi^+_{1-n})^*&(e^{\mathbf{K}_{m+1}}\hat\Psi^+_{-n})^*& \cdots
        &(e^{(2m+1)\mathbf{K}_{m+1}}\hat\Psi^+_{-n})^*
        &\gamma_n(e^{\mathbf{K}_{m+1}}\hat\Phi^+_{1-n})^*
        &\cdots &\gamma_n(e^{(2m-1)\mathbf{K}_{m+1}}\hat\Phi^+_{1-n})^*
    \end{vmatrix},
\end{align*}
}
from which one can calculate $Q_n$ of this case by $Q_n=\hat G_n/\hat F_n$.

\section{Conclusions}\label{sec-6}

In this paper,  by means of the B-R approach,
we have solved the four classical and nonlocal sdNLS equations in \eqref{sec2-dNLS}
with nonzero backgrounds.
In the B-R approach, we first solved the unreduced AL-2 system \eqref{sec2-AL2},
presenting its bilinear form \eqref{sec3-bilinear-new} and
quasi double Casoratian solutions (see Theorem \ref{Th-1}).
Note that in the bilinear form the background solutions $(q_n, r_n)$ are involved.
The reduction step is to impose constraints on the Casoratian column vectors $\Phi_n$ and $\Psi_n$,
together with the constraint equations on the spectral matrix $A$ and transform matrix $T$
(see the column `constraint' in Table \ref{tab-1}),
such that the quasi double Casoratians with different settings (e.g. complex conjugate, reverse space, etc)
are connected to each other
(see the column `$F_n, G_n, H_n$' in Table \ref{tab-1}).
This then gives rise to various connections between $Q_n=G_n/F_n$ and $R_n=H_n/F_n$,
which brings solutions for the reduced sdNLS equations in \eqref{sec2-dNLS}.
After that, we presented explicit forms for the satisfied matrices $A, T$ (or $B, T$) and the vectors $\Phi_n$ and $\Psi_n$,
which finally give rise to explicit solutions $Q_n$ for the reduced sdNLS equations.
All these results allow degenerations to the zero background case when $(q_n,r_n)=(0,0)$.

The advantage of the B-R approach is apparent.
It first solves the unreduced equations. At this stage there is not any complex conjugate operation involved.
Then, the reduction step employs a technique to deal with complex conjugation and reverse-space, etc
in the reduction. This is more convenient than directly solve those nonlinear equations
involving with complex conjugation, e.g. the NLS equation \eqref{nls} and the sdNLS equation \eqref{sdnls}.
In addition, it is also difficult to solve nonlocal equations directly using bilinear method (cf.\cite{MZ-AML-2014}).
Instead, the B-R approach has proved convenient and effective
in solving nonlocal equations \cite{CDLZ-SAPM-2018,CZ-AML-2018,DLZ-AMC-2018,SSZ-ND-2019,CLZ-AML-2019,LWZ-ROMP-2020,WWZ-2020,
W-ND-2021,LWZ-ROMP-2022,LWZ-SAPM-2022,WW-CNSNS-2022,
WW-ND-2022,WWZ-CPB-2022}.
Apart from the B-R approach, the KP-reduction approach can also be used to find solutions for nonlocal equations,
e.g.\cite{HC-CPL-2018,CFJ-WM-2020}, but it is hard to classify solutions in  the KP-reduction approach.
Note that the B-R approach allows classification of solutions according to the canonical forms of the related spectral matrix.
It is also notable that in this paper we have got explicit quasi double Casoratian forms
for the general rational solutions (rogue waves) for the classical focusing sdNLS equation \eqref{sdnls},
and we also presented a bilinear form \eqref{sec3-bilinear-sdnls} for the sdNLS equation \eqref{sdnls},
which is different from the one obtained in \cite{OY-JPA-2014}.

In Sec.\ref{sec-5}, we only analyzed and illustrated some solutions which are not singular.
Note that the hyperbolic background solutions \eqref{qr-tanh}
admit reduction $q_n=r_n^*$,
but the resulted solutions for the classical defocusing sdNLS equation
seem either singular or trivial. We do not present them in this paper.

As for possible topics for further investigation, we mention the following.
First, a recent remarkable result on rogue waves is their patterns \cite{YY-PD-2021a,YY-PD-2021b},
which are related to zeros of some special polynomials.
The patterns of the rogue waves obtained in this paper will be analyzed elsewhere.
Second, we have obtained explicit solutions for the classical focusing sdNLS equation with
plane wave background. This equation is related the nonintegrable sdNLS equation \eqref{sdnls-n},
which can be studied as a perturbation of \eqref{sdnls} \cite{CKKS-PRB-1993,BK-PR-1998}.
In addition, the sdNLS equation \eqref{sdnls} is also connected with the Heisenberg lattice \cite{I-JNMP-1982}and
the Toda lattice \cite{AL-JMP-1976,APT-book-2004}.
The obtained solutions of the sdNLS equation with plane wave background
may be used to study the nonintegrable sdNLS equation, Heisenberg lattice and Toda lattice.
The third one  is about the sdNLS equation with elliptic function backgrounds.
The focusing sdNLS equation admits simple elliptic function solutions \cite{CCX-PLA-2006,HL-CSF-2009,CP-SAPM-2024}
and also rogue waves standing on an elliptic function background \cite{CP-SAPM-2024}
(cf. \cite{CPW-PD-2020,FLT-SAPM-2020} for the continuous focusing NLS equation).
Considering elliptic solitons are popular in integrable systems \cite{NSZ-CMP-2023}
and some bilinear technique are already developed \cite{LZ-JNS-2022},
it would be interesting to extend the B-R approach
to the sdNLS equation with elliptic function backgrounds.
Finally, the B-R approach has recently been applied to a fully discrete NLS equation
with zero background \cite{ZFF-arX-2024}
and rogue waves of the same equation has been obtained via the KP-reduction method \cite{OF-PD-2022}.
The  fully discrete NLS equation with nonzero backgrounds will be investigated later by means of the B-R approach,
which will bring a classification of solutions of the equation.

\vskip 20pt
\subsection*{Acknowledgments}
This project is supported by the NSFC grant (No. 12271334, 12201580).

\vskip 20pt
\appendix

\section{Proof of theorem \ref{Th-1}} \label{app1-sec1}

Before starting the proof,  we introduce an identity \cite{FN-PLA-1983}
\begin{eqnarray}\label{app1-possion}
|\mathbf{M},\mathbf{a},\mathbf{b}||\mathbf{M},\mathbf{c},\mathbf{d}|
-|\mathbf{M},\mathbf{a},\mathbf{c}||\mathbf{M},\mathbf{b},\mathbf{d}|
+|\mathbf{M},\mathbf{a},\mathbf{d}||\mathbf{M},\mathbf{b},\mathbf{c}|=0,
\end{eqnarray}
where $\mathbf{M}$ is a $s\times(s-2)$ matrix,
$\mathbf{a}, \mathbf{b}, \mathbf{c}$ and $\mathbf{d}$ are $s$-th order column vectors.

We also use the following shorthand
\[        |\caso{\beta,\beta+2\mu},{\zeta};\caso{\gamma,\gamma+2\nu},{\xi}|
        =|A^\beta\Phi_n,A^{\beta+2}\Phi_n,\cdots,A^{\beta+2\mu}\Phi_n,A^{\zeta}\Phi_n;
        A^\gamma\Phi_n,A^{\gamma+2}\Phi_n,\cdots,A^{\gamma+2\nu}\Psi_n,A^{\xi}\Psi_n|,
\]
with which the determinants $F_n, G_n$ and $H_n$ given in \eqref{sec3-transform33}
can be denoted as
\begin{subequations}
\begin{align}
        F_n&=|\caso{1,2m+1};\caso{0,2m}|,\\
        G_n&=|\caso{0,2m+2};\caso{1,2m-1}|+(-1)^mq_n|\caso{0,2m};\caso{1,2m+1}|,\\
        H_n&=|\caso{2,2m};\caso{-1,2m+1}|+(-1)^mr_n|\caso{0,2m};\caso{1,2m+1}|.
    \end{align}
\end{subequations}
Direct calculations yield
\begin{align*}
\alpha_n|A| F_{n+1}=\,&|\caso{3,2m+3};\caso{0,2m}|-(-1)^{m}q_n|\caso{3,2m+1};\caso{0,2m+2}|\\
    &-(-1)^mr_n|\caso{1,2m+3};\caso{2,2m}|-q_nr_n|\caso{1,2m+1};\caso{2,2m+2}|,\\
|A|F_{n-1}=\,&|\caso{1,2m+1};\caso{2,2m+2}|,\\
i|A|\partial_tG_{n}=\,&-(2+q_{n-1}r_n)|\caso{1,2m+3};\caso{2,2m}|+(-1)^miq_t|\caso{1,2m+1};\caso{2,2m+2}|\\
    &+\frac12|\caso{1,2m+1},2m+5;\caso{2,2m}|-\frac12|\caso{1,2m+3};{0},\caso{4,2m}|\\
    &+\frac12|-1,\caso{3,2m+3};\caso{2,2m}|-\frac12|\caso{1,2m+3};\caso{2,2m-2},2m+2|\\
    &+\frac12(-1)^{m}q_n|\caso{1,2m+1};\caso{2,2m},2m+4|-\frac12(-1)^mq_n|\caso{1,2m+1};0,\caso{4,2m+2}|\\
    &+\frac12(-1)^mq_n|-1,\caso{3,2m+1};\caso{2,2m+2}|-\frac12(-1)^mq_n|\caso{1,2m-1},2m+3;\caso{2,2m+2}|\\
    &+(-1)^mq_{n-1}|\caso{3,2m+3};\caso{0,2m}|-(-1)^mq_{n-1}|\caso{1,2m+1};\caso{2,2m+2}|\\
    &-q_nq_{n-1}|\caso{3,2m+1};\caso{0,2m+2}|,\\
i\partial_tF_n=\,&\frac12|\caso{1,2m-1},2m+3;\caso{0,2m}|
-\frac12|\caso{1,2m+1};-2,\caso{2,2m}|\\
    &+\frac12|-1,\caso{3,2m+1};\caso{0,2m}|-\frac12|\caso{1,2m+1};\caso{0,2m-2},2m+2|\\
    &-(-1)^mq_n|\caso{1,2m-1};\caso{0,2m+2}|+(-1)^mr_n|\caso{-1,2m+1};\caso{2,2m}|,\\
\alpha_n G_{n+1}=\,&|\caso{1,2m+3};\caso{0,2m-2}|+(-1)^mq_n|\caso{1,2m+1};\caso{0,2m-2},{2m+2}|\\
    &-(-1)^mq_n|\caso{1,2m-1},2m+3;\caso{0,2m}|+(-1)^m\alpha_nq_nF_n\\
    &+q_nq_n|\caso{1,2m-1};\caso{0,2m+2}|+(-1)^m\alpha_nq_{n+1}F_n,\\
G_{n-1}=\,&|\caso{-1,2m+1};\caso{2,2m}|+(-1)^mq_{n-1}F_n.
\end{align*}
Substituting them into  \eqref{sec3-bilinear-new-equ1}, one obtains
\begin{align*}
&|A|^2(\alpha_nF_{n+1}F_{n-1}+G_nH_n-F_nF_n)\\
=\,&|\caso{3,2m+3};\caso{0,2m}||\caso{1,2m+1};\caso{2,2m+2}|
+|\caso{1,2m+3};\caso{2,2m}||\caso{3,2m+1};\caso{0,2m+2}|\\
&-|\caso{3,2m+3};\caso{2,2m+2}|F_n,
\end{align*}
which vanishes in light of the identity \eqref{app1-possion}.
Equation \eqref{sec3-bilinear-new-equ2} yields
\[|A|(iD_tG_{n}\cdot F_n-\alpha_n(G_{n+1}F_{n-1}+G_{n-1}F_{n+1})+2G_nF_n)
=S_1+S_2+(-1)^m q_n (S_3+S_4),\]
where
\begin{align*}
S_1=\,&-|\caso{-1,2m+1};\caso{2,2m}||\caso{3,2m+3};\caso{0,2m}|
-\frac12|\caso{1,2m+3};0,\caso{4,2m}|F_n
+\frac12|-1,\caso{3,2m+3};\caso{2,2m}|F_n\\
&+\frac12|\caso{1,2m+1};-2,\caso{2,2m}||\caso{1,2m+3};\caso{2,2m}|
-\frac12|-1,\caso{3,2m+1};\caso{2,2m}||\caso{1,2m+3};\caso{2,2m}|,\\
S_2=\,& -|\caso{1,2m+3};\caso{0,2m-2}||\caso{1,2m+1};\caso{2,2m+2}|
+\frac12|\caso{1,2m+1},2m+5;\caso{2,2m}|F_n\\
& -\frac12|\caso{1,2m+3};\caso{2,2m-2},2m+2|F_n
 -\frac12|\caso{1,2m-1},2m+3;\caso{0,2m}||\caso{1,2m+3},\caso{2,2m}|\\
&+\frac12|\caso{1,2m+1};\caso{0,2m-2},2m+2||\caso{1,2m+3};\caso{2,2m}|,\\
S_3=\, &|\caso{-1,2m+1};\caso{2,2m}||\caso{3,2m+1};\caso{0,2m+2}|
-\frac12|\caso{1,2m+1};0,\caso{4,2m+2}|F_n\\
& +\frac12|-1,\caso{3,2m+1};\caso{2,2m+2}|F_n
 +\frac12|\caso{1,2m+1};-2,\caso{2,2m}||\caso{1,2m+1};\caso{2,2m+2}|\\
& -\frac12|-1,\caso{3,2m+1};\caso{0,2m}||\caso{1,2m+1};\caso{2,2m+2}|,\\
S_4=\, & +|\caso{1,2m-1};\caso{0,2m+2}||\caso{1,2m+3};\caso{2,2m}|
+\frac12|\caso{1,2m+1};\caso{2,2m},2m+4|F_n\\
& -\frac12|\caso{1,2m-1},2m+3;\caso{2,2m+2}|F_n
 -\frac12|\caso{1,2m+1};\caso{0,2m-2},2m+2||\caso{1,2m+1};\caso{2,2m+2}|\\
& +\frac12|\caso{1,2m-1},2m+3;\caso{0,2m}||\caso{1,2m+1};\caso{2,2m+2}|.
\end{align*}
Each $S_j$ vanishes by using the identity \eqref{app1-possion} twice.
Thus, \eqref{sec3-bilinear-new-equ2} is verified.
Equation \eqref{sec3-bilinear-new-equ3} can be proved in a similar way.

Suppose $\Gamma$ is a matrix which is similar to $A$ via
 a transformation matrix $P$, i.e. $A=P^{-1}\Gamma P$.
We can introduce $\Phi'_n=P\Phi_n$ and $\Psi'_n=P\Psi_n$,
which again satisfy matrix equations \eqref{sec3-phipsi} with matrix $A$ replaced by $\Gamma$.
The quasi double Casoratians yield
\begin{align*}
& F_n(\Gamma, \Phi'_n,\Psi'_n)=|P|F_n(A,\Phi_n,\Psi_n), ~~
G_n(\Gamma,\Phi'_n,\Psi'_n)=|P|G_n(A,\Phi_n,\Psi_n),~~\\
& H_n(\Gamma,\Phi'_n,\Psi'_n)=|P|H_n(A,\Phi_n,\Psi_n),
\end{align*}
which indicates that  $A$ and $\Gamma$ lead to same $Q_n$ and $R_n$.

We complete the proof for Theorem \ref{Th-1}.

\end{document}